\documentclass[%
reprint,
showpacs,
amsmath,amssymb,
aps,
prb,
]{revtex4-1}

\usepackage{graphicx}% Include figure files
\usepackage{dcolumn}% Align table columns on decimal point
\usepackage{bm}% bold math
\usepackage{graphicx}% Include figure files
\usepackage{dcolumn}% Align table columns on decimal point
\usepackage{bm}% bold math
\usepackage{bigints}
\usepackage{color}
\usepackage{amsmath,amssymb,graphicx}
\usepackage{float}
\usepackage{braket}

\begin{document}

\preprint{APS/123-QED}

\title{Scaling analysis of transverse Anderson localization in a disordered optical waveguide}

\author{Behnam Abaie}
\author{Arash Mafi}%
 \email{mafi@unm.edu}
\affiliation{Department of Physics \& Astronomy, University of New Mexico, Albuquerque, NM 87131, USA
             \\Center for High Technology Materials, University of New Mexico, Albuquerque, NM 87106, USA}%

\date{\today}% It is always \today, today,

\begin{abstract}
The intention of this manuscript is twofold. First, the mode-width probability density function (PDF) is introduced as a powerful
statistical tool to study and compare the transverse Anderson localization properties of a disordered one dimensional optical waveguide.
Second, by analyzing the scaling properties of the mode-width PDF with the transverse size of the waveguide, it is shown that the mode-width PDF
gradually converges to a terminal configuration. Therefore, it may not be necessary to study a real-sized disordered structure 
in order to obtain its statistical localization properties and the same PDF can be obtained for a substantially smaller structure.
This observation is important because it can reduce the often demanding computational effort that is required to study the statistical properties 
of Anderson localization in disordered waveguides. Using the mode-width PDF, substantial information about the impact of the waveguide parameters
on its localization properties is extracted. This information is generally obscured when disordered waveguides are analyzed using other 
techniques such as the beam propagation method. As an example of the utility of the mode-width PDF, it is shown that the cladding refractive index can be used to
quench the number of extended modes, hence improving the contrast in image transport properties of disordered waveguides. 
%\begin{description}
%\item[]
%\end{description}
\end{abstract}
\maketitle
%%%%%%%%%%%%%%%%%%%%%%%%%%%%%%%%%%%%%%%%%%%%%%%%%%%%%%%%%%%%%%%%%%%%%%%
%%%%%%%%%%%%%%%%%%%%%%%%%%%%%%%%%%%%%%%%%%%%%%%%%%%%%%%%%%%%%%%%%%%%%%%
\section{Introduction}
%%%%%%%%%%%%%%%%%%%%%%%%%%%%%%%%%%%%%%%%%%%%%%%%%%%%%%%%%%%%%%%%%%%%%%%
%%%%%%%%%%%%%%%%%%%%%%%%%%%%%%%%%%%%%%%%%%%%%%%%%%%%%%%%%%%%%%%%%%%%%%%
Anderson localization has been a topic of great scientific interest for over five decades~\cite{Anderson1,EAbrahams,SegevNaturePhotonicsReview}.
It has been successfully demonstrated in highly scattering classical wave systems including acoustics~\cite{Weaver,Graham},
electromagnetics~\cite{Dalichaouch}, optics~\cite{John1,John2,John3,Anderson2,Lagendijk1,Hu,Chabanov}, as well as 
quantum optical systems, such as atomic lattices~\cite{Billy} and propagating photons~\cite{Lahini2,Lahini3,Abouraddy,Thompson}.
Transverse Anderson localization of light was first suggested by Abdullaev, et al.~\cite{Abdullaev} and De Raedt, et al.~\cite{DeRaedt}
and was confirmed experimentally by Schwartz, et al.~\cite{Schwartz}. In particular, De Raedt, et al. analyzed an optical 
waveguide with a transversely random and longitudinally invariant refractive index profile. They showed in this
quasi-two-dimensional (quasi-2D) system that an optical beam can propagate freely in the longitudinal direction while being trapped 
(Anderson localized) in the transverse direction. Transverse Anderson localization of light has since been observed in 
various quasi-one-dimensional (quasi-1D) and quasi-2D optical systems~\cite{Lahini,Martin,SalmanOL,SalmanOPEX,SalmanOMEX,MafiAOP}.

Most recently, Karbasi, et al. reported the first observation of transverse Anderson localization in disordered optical 
fibers~\cite{SalmanOL,SalmanOPEX,SalmanOMEX}. The disordered optical fibers were used for image transport~\cite{SalmanNature}
and it was shown that the high quality image transport was achieved because of, not in spite of, the high level of disorder 
and randomness in the fiber~\cite{SalmanImage,SalmanNature}. A high-quality imaging optical fiber based on transverse Anderson localization 
requires a {\em narrow} and {\em uniform} point spread function (PSF) across the tip of the fiber. The width of the PSF is determined 
by the localization radius; it has been argued that a large refractive index contrast is essential in ensuring that the localization 
radius is {\em sufficiently small} and {\em does not vary appreciably} across the fiber profile~\cite{SalmanOPEX,SalmanOMEX,SalmanNature}. 
The large refractive index fluctuations of aorund 0.1 in the disordered polymer fiber by Karbasi, et al. ensured a {\em strong} and 
{\em uniform} transverse localization across the tip of the fiber for high quality image transport.

It has been previously shown that coherent waves in one-dimensional (1D) and and two-dimensional (2D) unbounded disordered systems
are always localized~\cite{Abrahams}. For bounded 1D and 2D systems, if the sample size is considerably larger than the
localization radius, the boundary effects are minimal and can often be ignored~\cite{Szameit2010,Jovic,BoundaryPaper}.
A disordered waveguide can support both localized and extended modes simultaneously because it is transversely bounded.
By the extended, we refer to those modes which span the transverse dimension(s) of the waveguide whose transverse size 
also increases with increasing the transverse dimension(s) of the waveguide. However, as we shall see later, the occurrence 
of these modes becomes less probable as the transverse size of the waveguide becomes larger; and becomes improbable
when the disordered waveguide becomes sufficiently wide to be approximated by an unbounded disordered system. 

Because the refractive index of a disordered waveguide is random, the properties of the localized and extended modes 
should be studied using statistical techniques. The 
most relevant physical quantity that characterizes the localization properties of disordered waveguides is the mode width,
which is defined rigorously in Eq.~\ref{eq:sigma} and characterizes the transverse size of a guided mode. 
Therefore, detailed understanding of the {\em mode-width statistics} is the gateway to uncovering the localization properties.
In this manuscript, we report on the {\em probability density function} (PDF) of the mode width as a powerful tool 
to study Anderson localization in disordered waveguides. Using the {\em mode-width PDF}: we can obtain the average width of the 
localized modes which determines the size of the PSF for a disordered imaging fiber; we can obtain the standard-deviation of the mode-width 
distribution which determines the uniformity of the transported image across the tip of the fiber; we can obtain the distribution 
of the extended modes which affect the image contrast; and we can study the impact of the total size of the structure and the
cladding index contrast on the localized and extended modes.

The mode-width PDF contains all the relevant information on the mode-width statistics of a disordered waveguide ensemble. 
However, computing the mode-width PDF is a challenging numerical problem. In order to compute the PDF, a large number 
of different waveguide samples are generated in a given ensemble for proper statistics. 
The guided modes for each waveguide are calculated and the corresponding mode-width values are extracted to generate the PDF. 
Calculating the guided modes of even a single fiber structure may become highly challenging. For example,  
the V-number of the disordered polymer fiber in Ref.~\cite{SalmanOL} with air cladding is approximately 2,200 at 405~nm wavelength 
resulting in more than 2.3 million guided modes~\cite{SalmanOL}. 
Recall that the V-number is given by
%%%%%%%%%%%%
\begin{align}
V=\dfrac{\pi t}{\lambda}\sqrt{n^2_{\rm co}-n^2_{\rm cl}},
\label{eq:V-number}
\end{align}
%%%%%%%%%%%%
where $\lambda$ is the
optical wavelength, $t$ is the core diameter of the fiber (or the core width for the case of a 1D slab waveguide), and $n_{\rm co}$ 
($n_{\rm cl}$) is the effective refractive index of the core (cladding). The total number of the bound guided modes in a step-index 
optical fiber is $\approx V^2/2$.

Solving for all the guided modes for a given fiber and obtaining proper 
statistical averages over many fiber samples is a formidable task even for large computer clusters. However, a mode-width PDF is 
an absolutely necessary tool if one wants to obtain proper understanding of the behavior of the localized modes as well as the extended 
modes and their interaction with the fiber boundary.
In this manuscript, {\em we employ the power of scaling analysis} and study the mode-width PDF as a function of the total size of the 
waveguide. We show that the PDF converges to a terminal form as the waveguide dimensions are increased. As such, the mode-width PDF
of a real-sized disordered waveguide may be obtained by simulating a waveguide ensemble of a considerably smaller dimensions. 
Moreover, we will show that the region of the PDF corresponding to the Anderson localized modes converges to its terminal form 
considerably faster than the entire PDF as the size of the structure grows larger, while the region of the PDF corresponding 
to the extended modes is rather generic looking.
Therefore, to obtain the most useful information corresponding to the Anderson-localized region of the PDF, it is often possible to
even further reduce the size of the structures resulting in substantial reduction in computational effort. In certain 
systems, this may actually turn the computational problem from nearly impossible to one that can be handled by moderate sized 
computer clusters.   
  
The focus of this manuscript is to establish a framework for a comprehensive analysis of the mode-width statistics for transverse Anderson localization in 
optical fibers in the future. 
%The main objective of such an analysis is to obtain proper understanding of the statistical distribution 
%of the width of the guided modes in an Anderson localized disordered optical fiber. 
A mode-width distribution that is more strongly 
peaked at narrower mode-width values is favored because it can result in a smaller PSF for imaging applications. 
Moreover, a narrower distribution of the mode-widths indicates PSF uniformity across the fiber. 
In order to lay the groundwork for understanding the scaling behavior of the statistical distribution for both the 
localized and extended modes in a 2D Anderson localizing optical fiber,
we have decided to present a comprehensive characterization of a 1D Anderson localized optical waveguide in this manuscript. This exercise is quite illuminating
as it sheds light on the scaling behavior of the statistical distribution of the mode widths and shows the extent of information that can 
be extracted from such an exercise. The detailed analysis of a 2D disordered fiber structure will be presented in a future publication.

Here, we have chosen to calculate the transverse electric (TE) guided modes of the disordered 
waveguide using finite element method 
(FEM) presented in Refs.~\cite{BoundaryPaperFIO, BoundaryPaper, El-Dardiry, Kartashov, Lenahan}. 
Similar observations can be drawn for 
transverse magnetic (TM) guided modes, but we limit our analysis to TE in this paper for simplicity. 
The appropriate partial differential equation that will be solved in this manuscript is 
the paraxial approximation to the Helmholtz equation for electromagnetic wave propagation 
in dielectrics
%%%%%%%%%%%%
\begin{align}
\label{eq:helmholtz}
\nabla^2_{\rm T}A({\bf x}_{\rm T})+n^2({\bf x}_{\rm T})k_0^2A({\bf x}_{\rm T})=\beta^2A({\bf x}_{\rm T})
\end{align}
%%%%%%%%%%%%
where $A({\bf x}_{\rm T})$ is the transverse profile of the (TE) electric field 
$E({\bf x}_{\rm T},z,t)=A({\bf x}_{\rm T})\exp(i \beta z-i\omega_0 t)$, $\beta$ is the propagation constant,
$n({\bf x}_{\rm T})$ is the (random) refractive index of the waveguide, ${\bf x}_{\rm T}$ is the one (two) transverse dimension(s)
in 1D (2D), $\omega_0=ck_0$, and $k_0=2\pi/\lambda$ where $c$ is the speed of light in vacuum. Equation~\ref{eq:helmholtz} is an eigenvalue problem in $\beta^2$ and guided modes are those
solutions (eigenfunctions) with $\beta^2>n^2_{\rm cl}k_0^2$.
For each guided mode, the mode width is defined as the standard deviation $\sigma$ of the (1D) normalized intensity 
distribution $I(x)\propto |A(x)|^2$ of the mode according to  
%%%%%%%%%%%%
\begin{align}
\label{eq:sigma}
	\sigma = \left(\int_{-\infty}^{+\infty}~(x-\bar{x})^2~I(x)~dx\right)^{1/2},
\end{align}
%%%%%%%%%%%%
where the mode center is defined as 
%%%%%%%%%%%%
\begin{align}
\label{eq:xbar}
\bar{x} &= \int_{-\infty}^{+\infty} x~I(x)~dx.
\end{align}
%%%%%%%%%%%%
$x$ is the spatial coordinate across the width of the waveguide and the mode intensity profile is normalized such that $\int_{-\infty}^{+\infty} I(x) dx = 1$. 
$\sigma$ is a measure of width of the modes i.e. a larger $\sigma$ signifies a wider mode intensity profile distribution. 

Finally, we would like to contrast the power of the statistical simulation in the modal method with that of the finite difference beam propagation 
method (FD-BPM)~\cite{Huang} employed earlier in Refs.~\cite{Schwartz,SalmanOPEX}. When using the FD-BPM to analyze the Anderson localization
in optical waveguide, one is always worried about the extent to which the results are dependent on 
the shape and size of the input beam. The modal description is superior because it relies solely on the physics of the disordered system and is 
independent of the properties of the external excitation~\cite{SalmanModal}. As it will be shown in this manuscript, the mode-width PDF can reveal 
the subtle interplay between Anderson localization, step-index waveguiding, and support of non-localized extended modes, all of which can be present 
simultaneously in a disordered waveguide--something that cannot be achieved using the FD-BPM analysis.

%%%%%%%%%%%%%%%%%%%%%%%%%%%%%%%%%%%%%%%%%%%%%%%%%%%%%%%%%%%%%%%%%%%%%%%
%%%%%%%%%%%%%%%%%%%%%%%%%%%%%%%%%%%%%%%%%%%%%%%%%%%%%%%%%%%%%%%%%%%%%%%
\section{1D disordered lattice index profile}
\begin{figure}[t]
\centerline{
\includegraphics[width=\columnwidth]{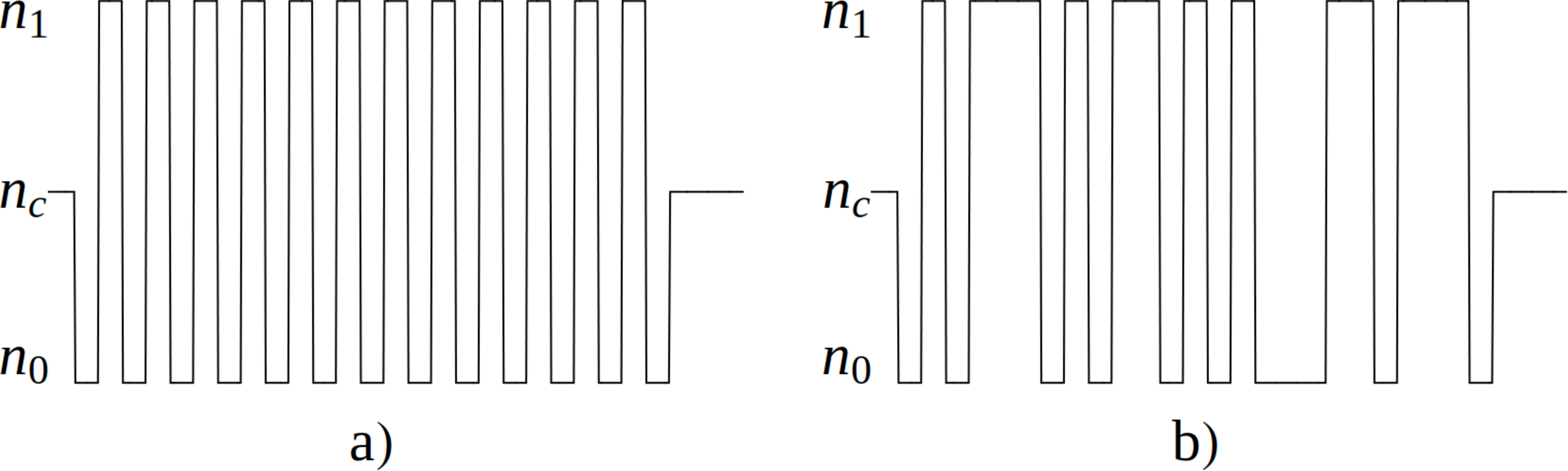}
}
\caption{\label{fig:IndexProfile}
Sample refractive index profiles of (a) ordered and (b) disordered slab waveguides are shown.}
\end{figure}

%%%%%%%%%%%%%%%%%%%%%%%%%%%%%%%%%%%%%%%%%%%%%%%%%%%%%%%%%%%%%%%%%%%%%%%
%%%%%%%%%%%%%%%%%%%%%%%%%%%%%%%%%%%%%%%%%%%%%%%%%%%%%%%%%%%%%%%%%%%%%%%
A 1D ordered optical lattice can be realized by periodically stacking dielectric layers with different refractive indexes 
on top of each other. Fig.~\ref{fig:IndexProfile}(a) shows the refractive index profile of a periodic 
1D optical waveguide where $n_{0}$, $n_{1}$, and $n_{c}$ correspond to the lower index layers, 
higher index layers, and cladding, respectively. In order to make a disordered waveguide, randomness can be introduced in different ways in 
the geometry or refractive index profile of a waveguide structure. For example, in 
Refs.~\cite{SalmanModal,BoundaryPaper} the thickness of the layers is randomized around an 
average value. In this manuscript, we adopt a different randomization method: we keep the thickness of all dielectric
layers identical but assign a refractive index value of $n_{0}$ or $n_{1}$ to each layer with a 50\% 
probability. This is the same as the method prescribed by De Raedt, et al.~\cite{DeRaedt} in randomizing a disordered 2D waveguide
which was also adopted by Karbasi, et al.~\cite{SalmanOL} to fabricate an Anderson localizing fiber. As we explained in
the Introduction, our intention is to extend our current analysis to 2D disordered Anderson localizing fibers in the future and 
we would like to stay as close as possible to the practical disordered 2D structure for proper comparison. 
%%%%%%%%%%%%%%%%%%%%%%%%%
%%%%%%%%%%%%%%%%%%%%%%%%%
%%%%%%%%%%%%%%%%%%%%%%%%%
%%%%%%%%%%%%%%%%%%%%%%%%%

\begin{figure}[t]
\centerline{
\includegraphics[width=\columnwidth]{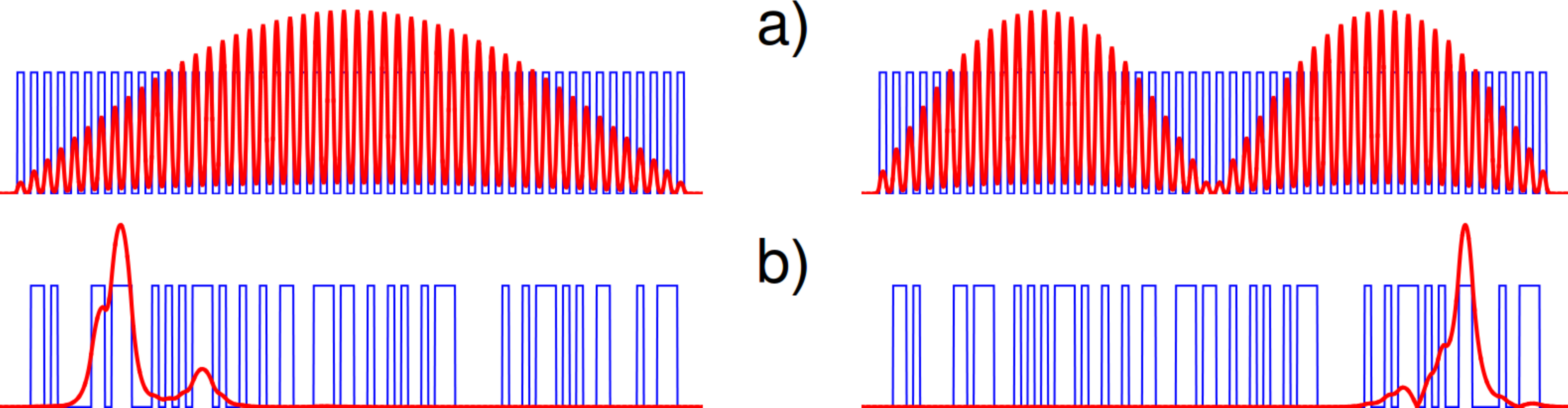}
}
\caption{\label{fig:ModeProfile}Typical mode profiles for (a) an ordered slab waveguide where each mode extends over 
the entire waveguide, and (b) a disordered slab waveguide, where the modes are localized.}
\end{figure}

Fig.~\ref{fig:IndexProfile}(b) shows the refractive index profile of a disordered 1D optical waveguide.
In Fig.~\ref{fig:ModeProfile}(a), we plot two guided modes of a 1D periodic waveguide with the 
highest propagation constant, where we have assumed that $n_{0}=1.49$, $n_{1}=1.50$, and $n_{c}=1.49$. 
These two modes belong to a large group of standard {\em extended} Bloch periodic guided modes supported 
by the ordered optical waveguide, which are modulated by the overall mode profile of the 1D waveguide~\cite{SalmanModal}.
The total number of guided modes depends on the total thickness and the refractive
index values of the slabs and cladding. The key point is that each mode of the periodic structure extends over the entire 
width of the waveguide structure. A similar exercise can be done with a 1D disordered waveguide, where two arbitrarily selected
modes are plotted in Fig.~\ref{fig:ModeProfile}(b) using the same refractive index parameters as that of the periodic waveguide. 
It is clear that the modes become localized in the disordered 1D waveguide. While there are variations in
the shape and width of the modes, the mode profiles shown in Fig.~\ref{fig:ModeProfile}(b) are typical.
%%%%%%%%%%%%%%%%%%%%%%%%%
%%%%%%%%%%%%%%%%%%%%%%%%%
%%%%%%%%%%%%%%%%%%%%%%%%%
%%%%%%%%%%%%%%%%%%%%%%%%%

It is important to note that the disordered core of the lattice is sandwiched between a cladding with a refractive index of $n_{c}$ 
that can also be adjusted to resemble experimental situations where a waveguide is surrounded by air or a dielectric with a refractive 
index higher than air or even $n_{0}$ (but always less than $n_{1}$ to ensure waveguiding). As we will see later, the value of $n_{c}$
influences the mode-width distribution of the extended modes in a 1D Anderson localized waveguide and should be carefully studied
in practical implementations of such structures, e.g. for image transport~\cite{SalmanNature}.

%%%%%%%%%%%%%%%%%%%%%%%%%%%%%%%%%%%%%%%%%%%%%%%%%%%%%%%%%%%%%%%%%%%%%%%
%%%%%%%%%%%%%%%%%%%%%%%%%%%%%%%%%%%%%%%%%%%%%%%%%%%%%%%%%%%%%%%%%%%%%%%
\section{Analysis of the mode-width PDF}
\begin{figure}[t]
\centerline{
\includegraphics[width=1.0\columnwidth]{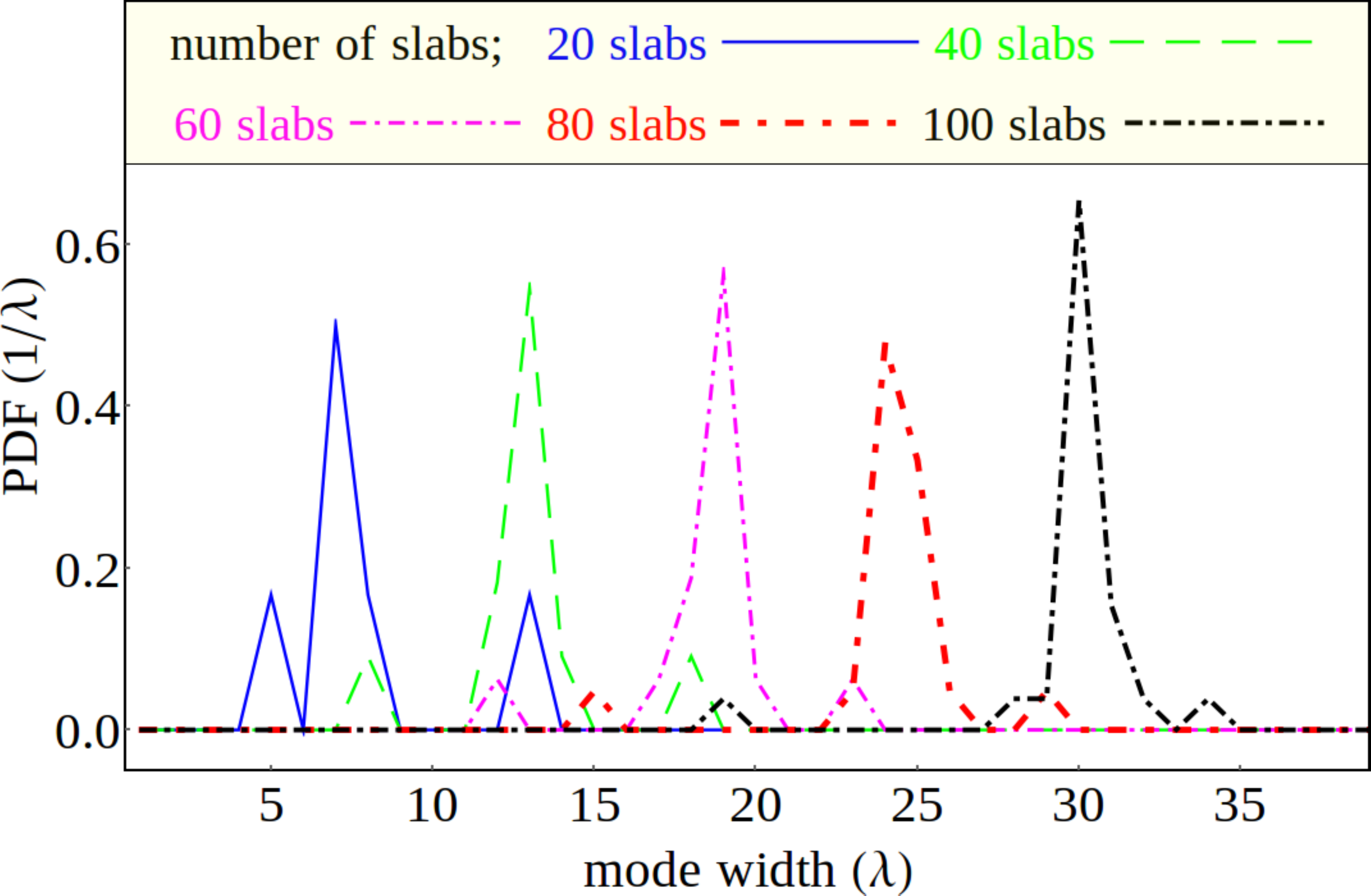}
}
\caption{\label{fig:PD2}Mode-width PDF of an ordered waveguide defined as ${\Delta n_{\rm core}} = 0.01$ and ${\Delta n_{\rm clad}} = 0$, for N = 20, 40, 60, 80, and 100 slabs. Mode width increases as the number of slabs increases, so the average mode width scales proportional to the size of the structure.}
\end{figure}

%%%%%%%%%%%%%%%%%%%%%%%%%%%%%%%%%%%%%%%%%%%%%%%%%%%%%%%%%%%%%%%%%%%%%%%
%%%%%%%%%%%%%%%%%%%%%%%%%%%%%%%%%%%%%%%%%%%%%%%%%%%%%%%%%%%%%%%%%%%%%%%
In the absence of localization, guided modes are Bloch periodic and extend over the entire width of a waveguide as shown in Fig.~\ref{fig:ModeProfile}(a).
In this case, the confinement is merely due to the total internal reflection at the effective index step between the waveguide and the cladding. 
In Fig.~\ref{fig:PD2}, we plot the mode-width PDF for the periodic waveguide with $N$ slabs, for
$N=20$, 40, 60, 80, and 100. The width of each slab is equal to the wavelength $d=\lambda$, $n_{c} = n_{0}=1.49$, and $n_1=1.50$ so the index step 
${\Delta n_{\rm core}} = 0.01$ where ${\Delta n_{\rm core}} = n_{\rm 1} - n_{\rm 0}$. The horizontal axis is in units of $\lambda$ and the vertical axis is in units of $1/\lambda$ such that the PDF integrates to one (unit area under the PDF curve).
The width of the cladding is assumed to be $25\lambda$ on each side of the waveguide. 
The guided modes decay exponentially (in the transverse direction) in the cladding. We have verified that the $25\lambda$ 
cladding is sufficiently wide so that the exponential decay in the cladding combined with the Dirichlet boundary condition of vanishing 
mode profile imposed at the outer edge of each cladding properly approximate an infinite cladding.
Figure~\ref{fig:PD2} shows that for the periodic slab waveguide,
the mode widths are determined by the width of the waveguide as can be seen clearly in Fig.~\ref{fig:ModeProfile}(a). Therefore, the mode widths, on the 
average, scale linearly with the size of the waveguide structure and the peak of the PDF shifts to larger values of mode width as the waveguide becomes wider. 
%%%%%%%%%%%%%%%%%%%%%%%%%
%%%%%%%%%%%%%%%%%%%%%%%%%
%%%%%%%%%%%%%%%%%%%%%%%%%
%%%%%%%%%%%%%%%%%%%%%%%%%

%%%%%%%%%%%%%%%%%%%%%%%%%
%%%%%%%%%%%%%%%%%%%%%%%%%
\begin{figure}[t]
\centerline{
\includegraphics[width=1.0\columnwidth]{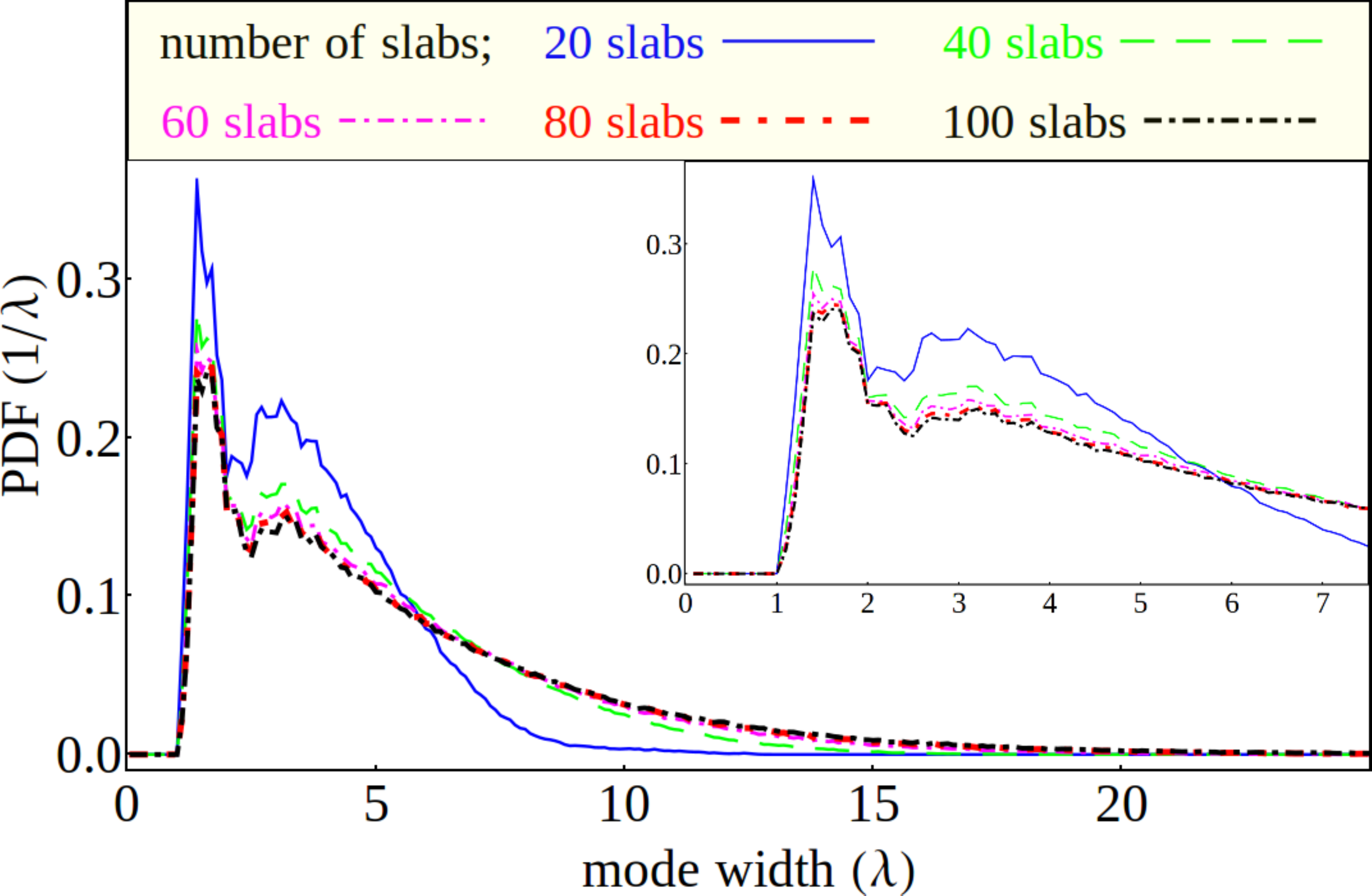}
}
\caption{\label{fig:PDFcase1} Mode-width PDF of a disordered waveguide defined by ${\Delta n_{\rm core}} = 0.01$, ${\Delta n_{\rm clad}} = 0$, 
for N = 20, 40, 60, 80, and 100 slabs. The inset is the magnified version of 
the localized peaks of the PDFs . Saturation of the PDF beyond $N_{\rm sat}\approx 60$ is clear in this figure.
}
\end{figure}
%%%%%%%%%%%%%%%%%%%%%%%%%
%%%%%%%%%%%%%%%%%%%%%%%%%
For a disordered waveguide, the scaling behavior of the PDF with the size of the waveguide is completely different from that of the periodic waveguide
shown in Fig.~\ref{fig:PD2}. When Anderson localization comes into play due to the disorder in the structure of waveguide, most guided modes become transversely 
localized as shown in Fig.~\ref{fig:ModeProfile}(b), while a few extended guided modes may still be supported depending on the waveguide configuration. 
As the number of slabs
is increased, the PDF saturates to a terminal form. In Fig.~\ref{fig:PDFcase1}, we show the PDF for an ensemble of disordered waveguides with $N$ slabs defined by
$n_{c} = n_{0}=1.49$, and $n_1=1.50$ (${\Delta n_{\rm core}} = 0.01$), where the width of each slab is equal to the wavelength $d=\lambda$. The PDFs are plotted for
$N=20$, 40, 60, 80, and 100. The PDF shows two localized peaks at width values less than $4\lambda$ with a long tail signifying the extended modes.  
The shape of the PDF changes with the number of slabs; however, it remains nearly unchanged beyond $N_{\rm sat}\approx 60$.
The near saturation of the PDF beyond a critical number of slabs $N_{\rm sat}$ is of utmost importance for two reasons: 1) $N_{\rm sat}$ can be viewed as the
effective transverse scale (waveguide width) beyond which the average localization dynamic is no longer dictated by the boundary; and 2) if we need to calculate the
PDF for a wide disordered waveguide, it is sufficient to simulate a waveguide with only $N_{\rm sat}$ slabs because it gives the same PDF; therefore, the 
computational effort can be significantly reduced. In order to see the saturation behavior of the PDF more clearly, the inset shows a magnified 
version of the PDFs, which is zoomed in at smaller mode width values. The transition to the terminal form of the PDF is clearly observed in the Anderson localized region of the PDFs where mode width is approximately less than $\approx 5\lambda$. 

%%%%%%%%%%%%%%%%%%%%%%%%%%%%%%%%%%%%%%%%%%%%%%%%%%%%%%%%%%%%%%%%%%%%%%%%%%%
\subsection{Impact of the index difference in the disordered waveguide}
%%%%%%%%%%%%%%%%%%%%%%%%%%%%%%%%%%%%%%%%%%%%%%%%%%%%%%%%%%%%%%%%%%%%%%%%%%%
%%%%%%%%%%%%%%%%%%%%%%%%%
%%%%%%%%%%%%%%%%%%%%%%%%%
\begin{figure}[t]
\centerline{
\includegraphics[width=1.0\columnwidth]{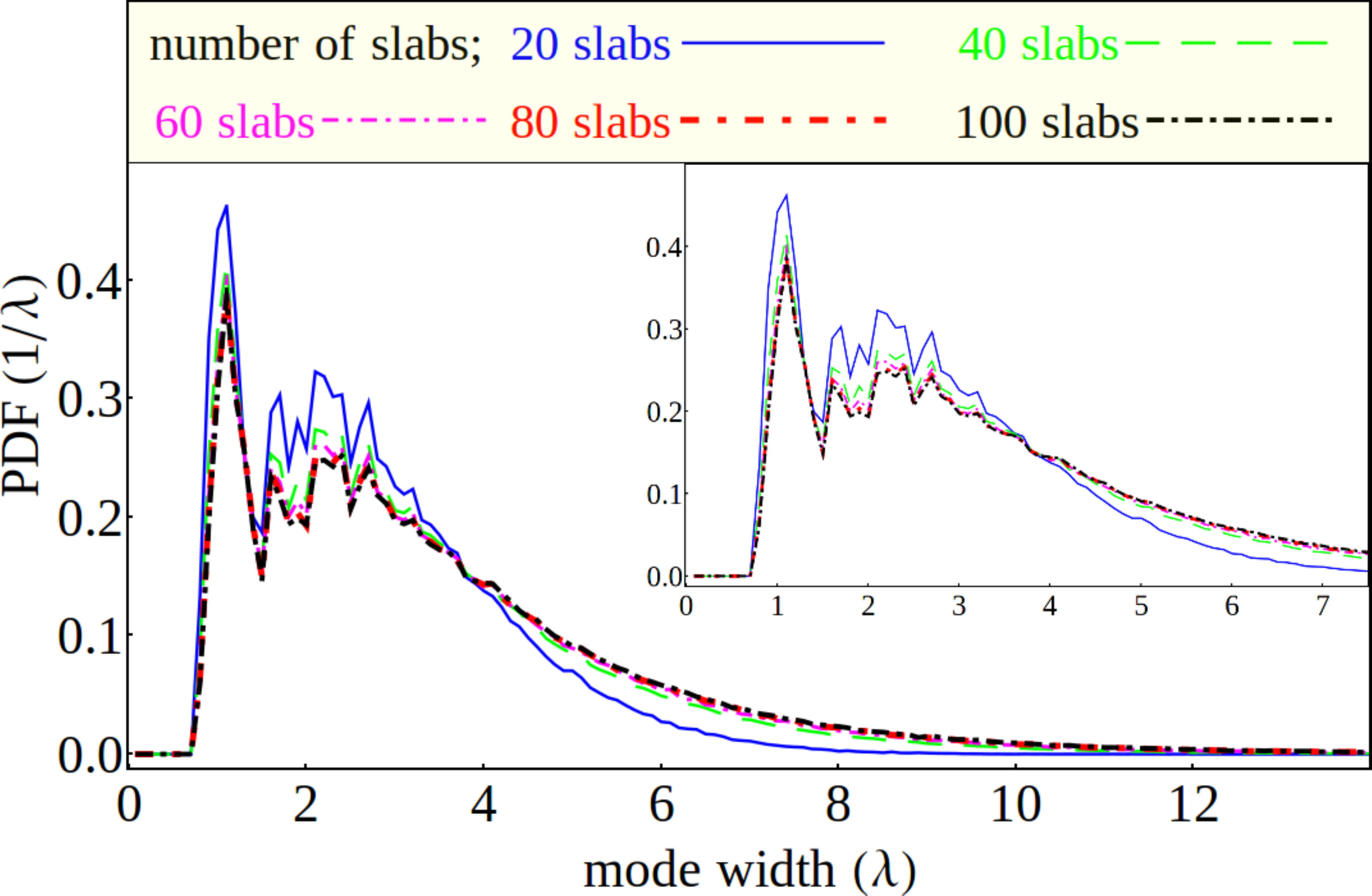}
}
\caption{\label{fig:PDFcase2}${\Delta n_{\rm core}} = 0.02$ is increased in comparison to Fig.~\ref{fig:PDFcase1}; The small-mode-width 
peak of the PDF shifts towards smaller mode width values indicating a stronger localization for a larger index contrast 
in the disordered waveguide. A magnified version is shown as an inset. $N_{\rm sat}\approx 40$ is smaller for a stronger transverse scattering.}
\end{figure}
%%%%%%%%%%%%%%%%%%%%%%%%%
%%%%%%%%%%%%%%%%%%%%%%%%%
The results shown in Fig.~\ref{fig:PDFcase1} are for the waveguide index difference of ${\Delta n_{\rm core}} = 0.01$ ($n_{0}=1.49$ and $n_1=1.50$).
If the index difference is increased, the stronger transverse scattering should result in stronger transverse localization and smaller localized mode 
width values. This can be seen by plotting the PDF for the higher index difference of ${\Delta n_{\rm core}} = 0.02$ ($n_{0}=1.48$ and $n_1=1.50$) 
in Fig.~\ref{fig:PDFcase2} and its magnified inset. The small mode width peak of the PDF relating to 
the Anderson localized modes 
in Fig.~\ref{fig:PDFcase2} has shifted to lower mode width values compared with Fig.~\ref{fig:PDFcase1} because of the larger $\Delta n$ and 
stronger transverse scattering.
Also, the convergence of the PDF happens with a smaller number of slabs, i.e., $N_{\rm sat}$ is smaller when $\Delta n$ is larger. 
Otherwise, the qualitative behavior of the PDFs are similar in the sense that the both waveguides support localized and extended modes simultaneously.  
%%%%%%%%%%%%%%%%%%%%%%%%%%%%%%%%%%%%%%%%%%%%%%%%%%%%%%%%%%%%%%%%%%%%%%%%%%%
\subsection{Impact of the boundary index difference}
%%%%%%%%%%%%%%%%%%%%%%%%%%%%%%%%%%%%%%%%%%%%%%%%%%%%%%%%%%%%%%%%%%%%%%%%%%%
In the previous figures (Figs.~\ref{fig:PDFcase1},~\ref{fig:PDFcase2}), the refractive index of the cladding $n_c$ 
is assumed to be the same as the refractive index $n_0$ of the lower index layers. For the practical 2D disordered optical fiber of Ref.~\cite{SalmanJOVE},
the cladding of the structure is air with a refractive index of $n_c=1$, which is considerably smaller than the lower index $n_{0}=1.49$ of the fiber.
The cladding index of the fiber can be controlled by an additional cladding layer or an index matching gel. As such, understanding the impact of the 
refractive index of the boundary on the guided mode structure of the disordered waveguide is of practical importance.  
A lower cladding index increases the effective V-number of the whole disordered waveguide, resulting in an increase in the total number of modes.
In this section, we will investigate the impact of the cladding refractive index on the distribution of the localized and extended modes, as well as 
on the scaling and eventual convergence of the mode-width PDF with the transverse size of the waveguide. Moreover, we will show that the impact of 
a change in the cladding index is primarily on the extended modes, while the localized modes are hardly affected by changes in the cladding index step. 

%%%%%%%%%%%%%%%%%%%%%%%%%
%%%%%%%%%%%%%%%%%%%%%%%%%
\begin{figure}[t]
\centerline{
\includegraphics[width=1.0\columnwidth]{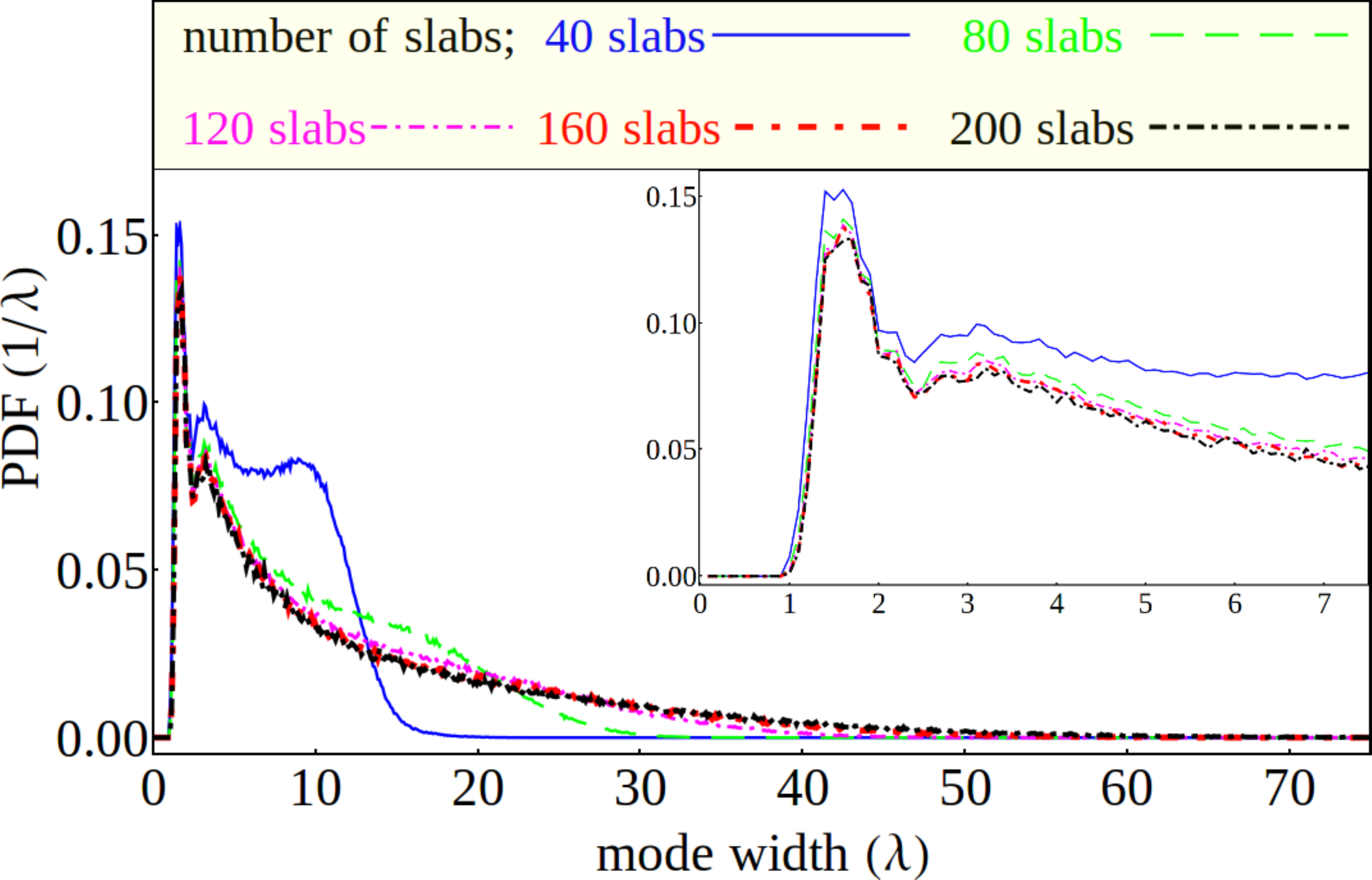}
}
\caption{\label{fig:PDFcase3}Mode-width PDF of a disordered waveguide defined as ${\Delta n_{\rm core}} = 0.01$, ${\Delta n_{\rm clad}} = 0.01$, and N = 40, 80, 120, 160, 200 slabs. A lower cladding index significantly changes the distribution of the extended modes and PDF saturates at a much larger value of N ($N_{\rm sat}\approx 160$). The inset represents a magnified version. A smaller cladding index only affects the extended mode width distribution, while the localized modes 
remain nearly unaffected.}
\end{figure}
%%%%%%%%%%%%%%%%%%%%%%%%%
%%%%%%%%%%%%%%%%%%%%%%%%%
In Fig.~\ref{fig:PDFcase3}, we consider a disordered waveguide with ${\Delta n_{\rm core}} = 0.01$ ($n_{0}=1.49$ and 
$n_1=1.50$) and ${\Delta n_{\rm clad}} = 0.01$ where  
${\Delta n_{\rm clad}} = n_{\rm 0} - n_{\rm c}$. This waveguide is identical in structure to that of Fig.~\ref{fig:PDFcase1} except 
for ${\Delta n_{\rm clad}}$. The main difference 
between Fig.~\ref{fig:PDFcase3} and Fig.~\ref{fig:PDFcase1} is in the distribution of the extended modes. The presence of a larger cladding index difference 
in Fig.~\ref{fig:PDFcase3} results in a greater number of extended modes which appears as a large bump in the PDF for $N=40$ slabs and smooths down 
when the PDF saturates to the terminal shape for large $N$. Another important difference between Fig.~\ref{fig:PDFcase3} and Fig.~\ref{fig:PDFcase1} is that
the convergence of the PDF in Fig.~\ref{fig:PDFcase3} (larger ${\Delta n_{\rm clad}}$) happens at a larger value of N.  
The inset in Fig.~\ref{fig:PDFcase3} (${\Delta n_{\rm clad}}=0.01$) shows the same PDF magnified over the region of small mode width near the localized modes
and should be compared with the inset in Fig.~\ref{fig:PDFcase1} 
(${\Delta n_{\rm clad}}=0$). While the
two figures are visually similar, the localized peak of Fig.~\ref{fig:PDFcase1} is observed to be clearly higher when comparing 
the vertical scales of the PDF plots. This is due to the fact that the total area under PDF is normalized to unity and the larger number of extended modes in 
Fig.~\ref{fig:PDFcase3} results in a reduction in the overall amplitude of the PDF over the entire domain. As such, the PDF in its present form
cannot provide a fair comparison between the localized mode structure of Fig.~\ref{fig:PDFcase3} and Fig.~\ref{fig:PDFcase1}. 
We will get back to this important point later in this section.  

\begin{figure}[t]
\centerline{
\includegraphics[width=1.0\columnwidth]{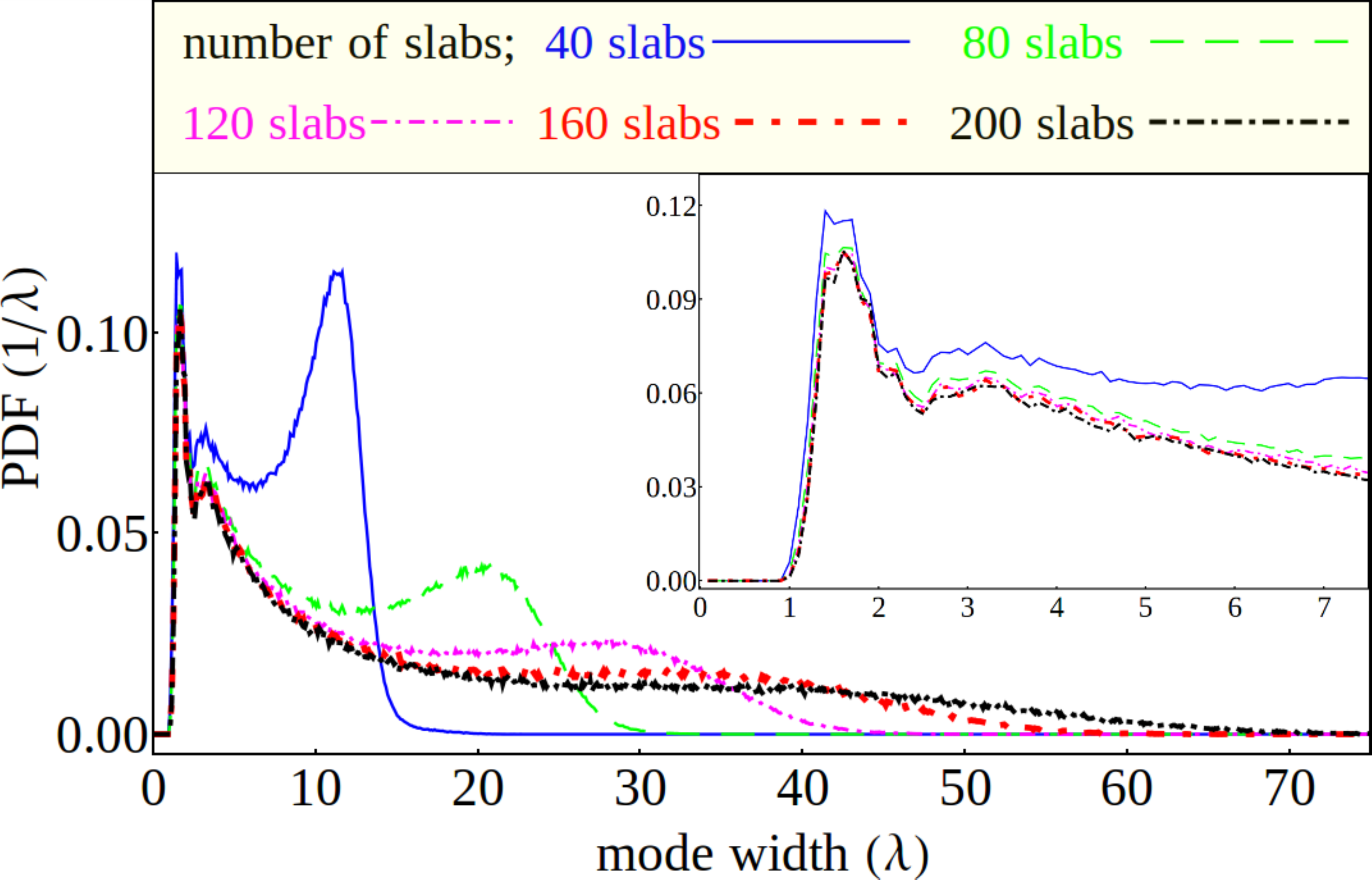}
}
\caption{\label{fig:PDFcase4}The same as Fig.~\ref{fig:PDFcase3} except ${\Delta n_{\rm clad}} = 0.02$. The PDF saturates to its terminal 
shape at a larger number of slabs ($N_{\rm sat}\approx 200$). The inset is a magnified version.}
\end{figure}

\begin{figure}[t]
\centerline{
\includegraphics[width=1.0\columnwidth]{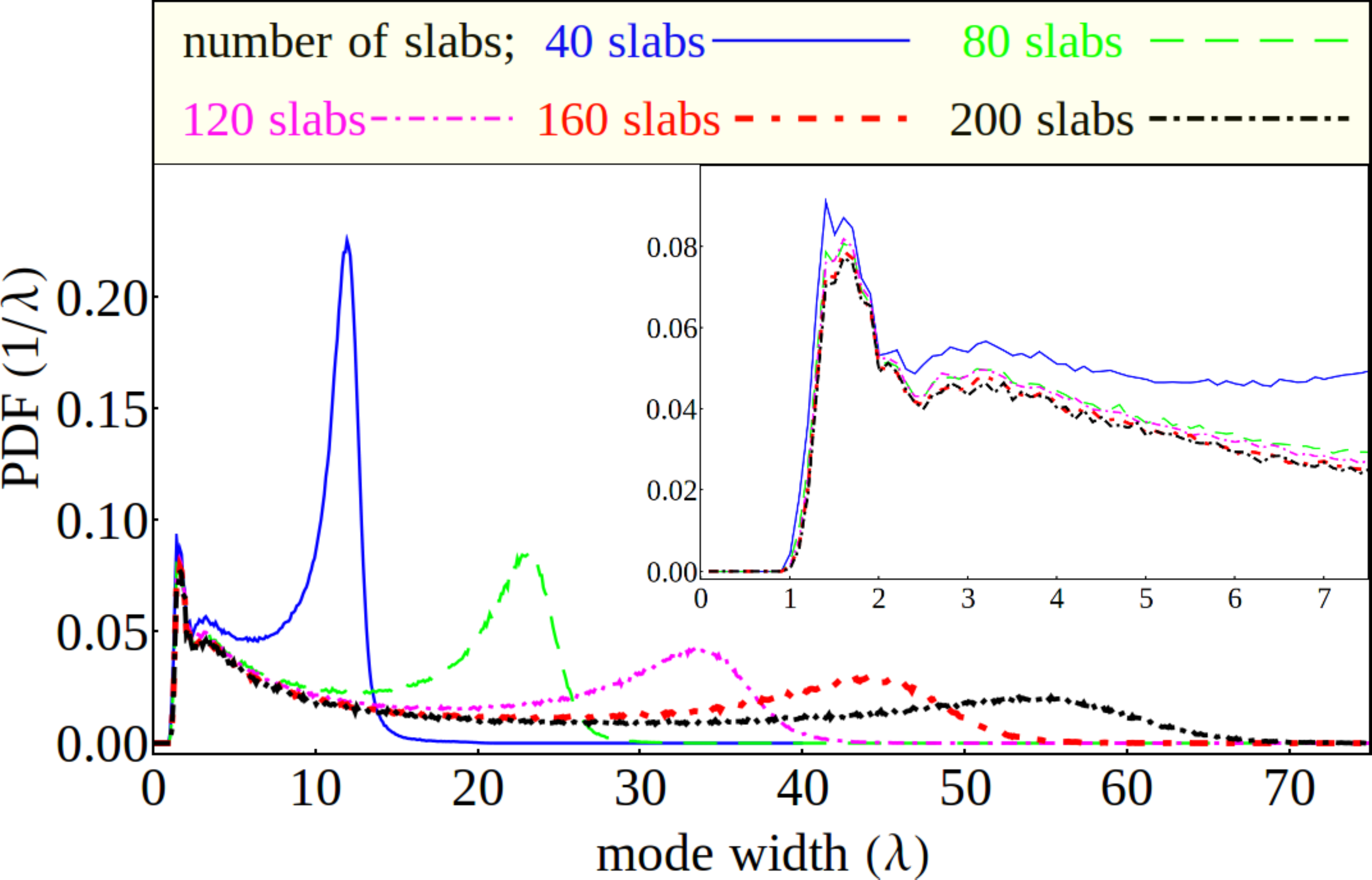}
}
\caption{\label{fig:PDFcase6}The same as Fig.~\ref{fig:PDFcase3} except ${\Delta n_{\rm clad}} = 0.04$. Further decreasing the cladding index delays the saturation of the PDF to larger values of N ($N_{\rm sat}\gg 200$). The inset is a magnified version.}
\end{figure}

In Fig.~\ref{fig:PDFcase4} and Fig.~\ref{fig:PDFcase6} we investigate the effect of further lowering $n_c$ to have 
${\Delta n_{\rm clad}} = 0.02$ and ${\Delta n_{\rm clad}} = 0.04$, respectively. The insets are the magnified 
versions as before over the region of small mode width near the localized modes. We observe a trend similar to the comparison that we 
conducted above between Fig.~\ref{fig:PDFcase3} and Fig.~\ref{fig:PDFcase1}. Therefore, we conclude that increasing 
${\Delta n_{\rm clad}}$ results in an increase in the number of extended modes, emphasizing that we have 
yet to show in this section that the localized modes 
are not affected by the change in the cladding index. Moreover, an increase in the cladding index difference results in a delayed 
convergence of the PDF to its terminal form resulting in a larger value of $N_{\rm sat}$. In fact, it can be seen that
$N_{\rm sat}\approx 200$ for Fig.~\ref{fig:PDFcase4} and $N_{\rm sat}\gg 200$ for Fig.~\ref{fig:PDFcase6}.  
%%%%%%%%%%%%%%%%%%%%%%%%%
%%%%%%%%%%%%%%%%%%%%%%%%%
%%%%%%%%%%%%%%%%%%%%%%%%%
%%%%%%%%%%%%%%%%%%%%%%%%%
%%%%%%%%%%%%%%%%%%%%%%%%%
%%%%%%%%%%%%%%%%%%%%%%%%%
%%%%%%%%%%%%%%%%%%%%%%%%%
%%%%%%%%%%%%%%%%%%%%%%%%%

\begin{figure}[t]
\centerline{
\includegraphics[width=1.0\columnwidth]{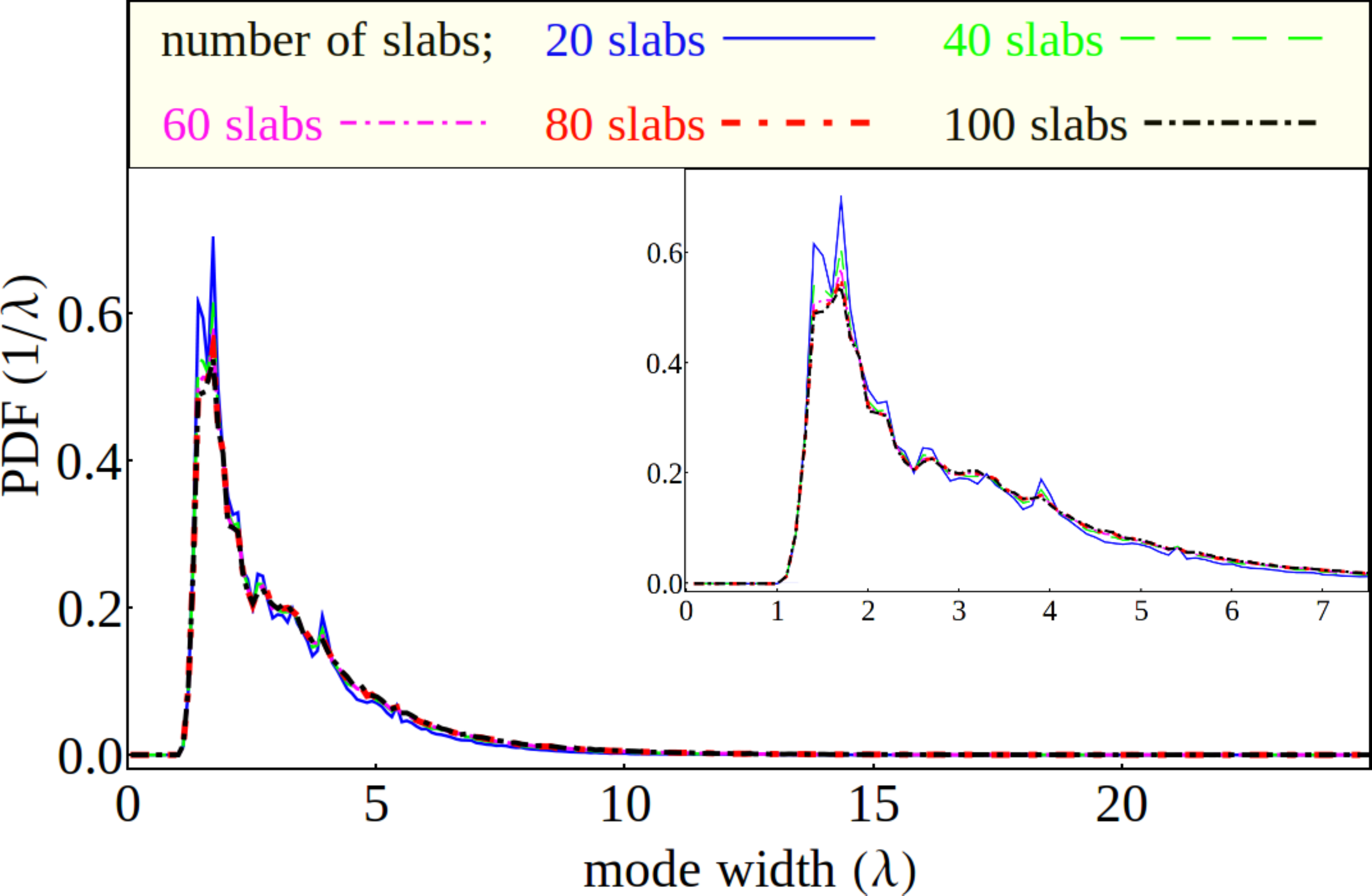}
}
\caption{\label{fig:PDFcase7}The same as Fig.~\ref{fig:PDFcase1} except ${\Delta n_{\rm clad}} = -0.005$. A cladding index 
larger than $n_{\rm 0}$ not only reduces the probability density of the extended modes, but also diminishes the second peak of the localized regime.
The inset represents a magnified version.}
\end{figure}

Our discussion will not be complete without discussing the reverse effect of raising $n_c$ above $n_0$, hence a negative
value of ${\Delta n_{\rm clad}} = -0.005$. Of course, we assume that $n_c<n_1$; otherwise, no guiding mode would exist.
Figure~\ref{fig:PDFcase7} (${\Delta n_{\rm clad}} = -0.005$) can best be compared with Fig.~\ref{fig:PDFcase1} with ${\Delta n_{\rm clad}} = 0$. 
It is clear that raising $n_c$ above $n_0$ (negative ${\Delta n_{\rm clad}}$) 
removes a considerable number of extended modes from the system. Recall that the PDF in Fig.~\ref{fig:PDFcase1}
showed two distinct peaks in the region near the localized modes and raising $n_c$ above $n_0$ appears to remove the second localized 
peak (with a larger mode width). Therefore, we conclude that a negative ${\Delta n_{\rm clad}}$ not only removes many of the
extended modes, it also removes those more weakly localized modes associated with the second peak in Fig.~\ref{fig:PDFcase1}. 
%%%%%%%%%%%%%%%%%%%%%%%%%
%%%%%%%%%%%%%%%%%%%%%%%%%
%%%%%%%%%%%%%%%%%%%%%%%%%
%%%%%%%%%%%%%%%%%%%%%%%%%

Previously in this Section, we mentioned that it is hard to judge the impact of the cladding refractive index on the width distribution of the
localized modes by comparing the PDFs from two different waveguides. The reason is that the total area under PDF is normalized to unity and 
different waveguide parameters result in different number of modes. As such, we need to come up with a method to clearly differentiate between 
the impact of the cladding index on the extended modes versus localized modes across different lattice parameters. In order to do this, it is best 
to use the {\em normalized PDF} which is the PDF multiplied by a constant factor such that total area under the normalized PDF curve 
equals the average number of modes in each class of random waveguide. 

\begin{figure}[t]
\centerline{
\includegraphics[width=1.0\columnwidth]{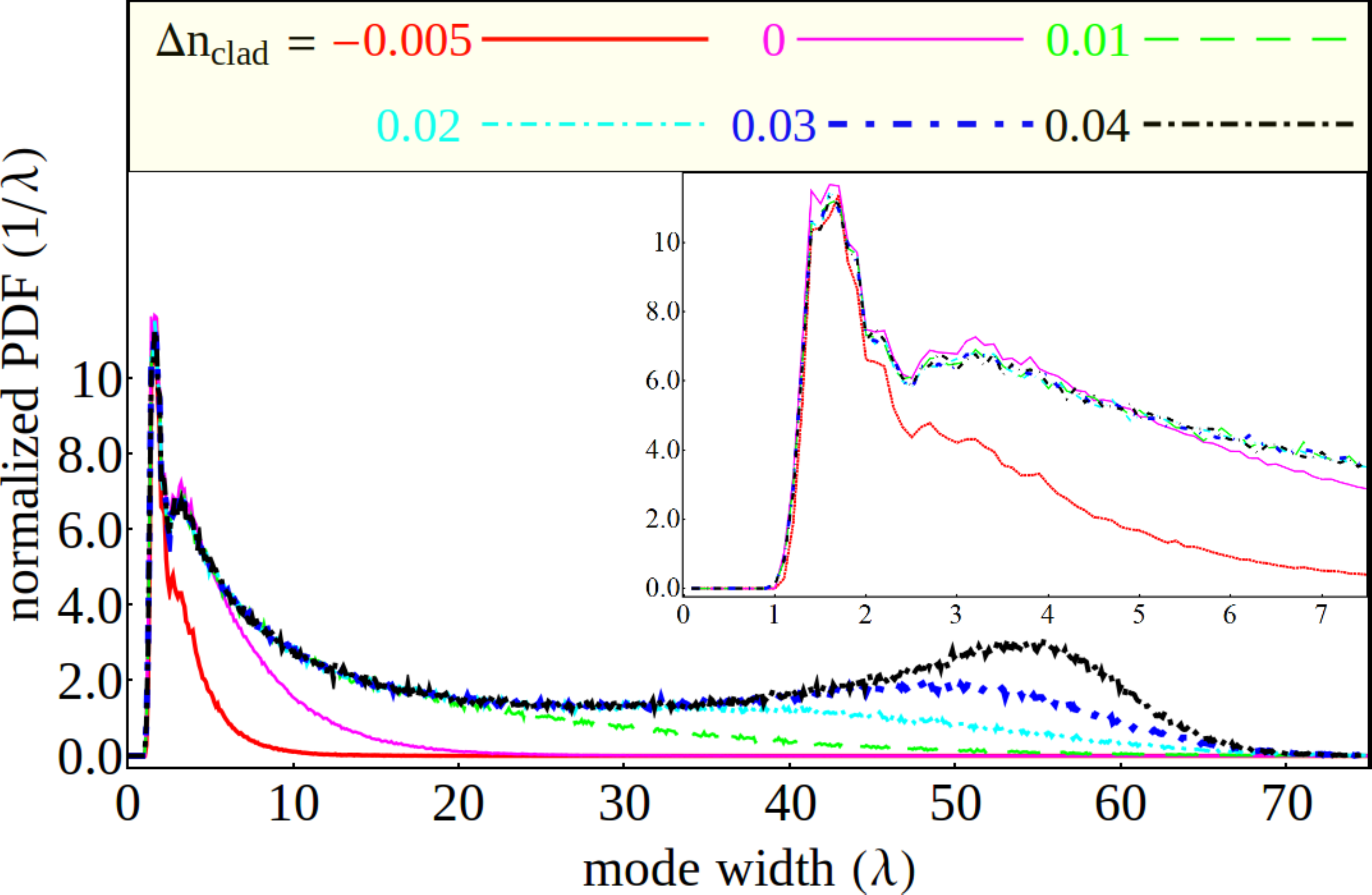}
}
\caption{\label{fig:PDFcase8}Normalized PDF for disordered waveguides defined by ${\Delta n_{\rm core}} = 0.01$, $N=200$, 
and ${\Delta n_{\rm clad}} =$ -0.005, 0.01, 0.02, 0.03, and 0.04. The magnified inset clearly shows that the statistics of the localized modes is independent of the cladding index unless for a negative ${\Delta n_{\rm clad}}$.}
\end{figure}

In Fig.~\ref{fig:PDFcase8}, we plot the normalized PDF for
disordered waveguides with ${\Delta n_{\rm core}} = 0.01$ ($n_{0}=1.49$ and $n_1=1.50$), $d=\lambda$, and $N=200$ slabs. Different curves
in Fig.~\ref{fig:PDFcase8} correspond to different values of ${\Delta n_{\rm clad}}$ ranging from -0.005 to 0.04. The curves belonging to the
largest three values of ${\Delta n_{\rm clad}}$ are not fully saturated to the terminal PDF because $N=200$ is smaller than $N_{\rm sat}$ in these cases,
hence resulting in a bump in the extend mode region. The inset shows the magnified version of the same figure in the region of
the localized modes. Figure~\ref{fig:PDFcase8} clearly shows that increasing the cladding index step merely introduces new extended modes and the localized 
modes are hardly affected. We re-emphasize the utility of the normalized PDF in revealing this important behavior. The case of
${\Delta n_{\rm clad}}=-0.005$ is quite interesting, as it can be seen that raising $n_c$ above $n_0$ strongly decouples extended modes and trims the
large-mode-width edge of the localized mode region of the PDF. Therefore, if having more localized modes versus extended modes is a desired outcome of 
a design, a small or even negative ${\Delta n_{\rm clad}}$ is preferable. 
%%%%%%%%%%%%%%%%%%%%%%%%%
%%%%%%%%%%%%%%%%%%%%%%%%%
%%%%%%%%%%%%%%%%%%%%%%%%%
%%%%%%%%%%%%%%%%%%%%%%%%%
%%%%%%%%%%%%%%%%%%%%%%%%%%%%%%%%%%%%%%%%%%%%%%%%%%%%%%%%%%%%%%%%%%%%%%%%%%%
\subsection{Impact of the unit slab thickness}
\begin{figure}[t]
\centerline{
\includegraphics[width=1.0\columnwidth]{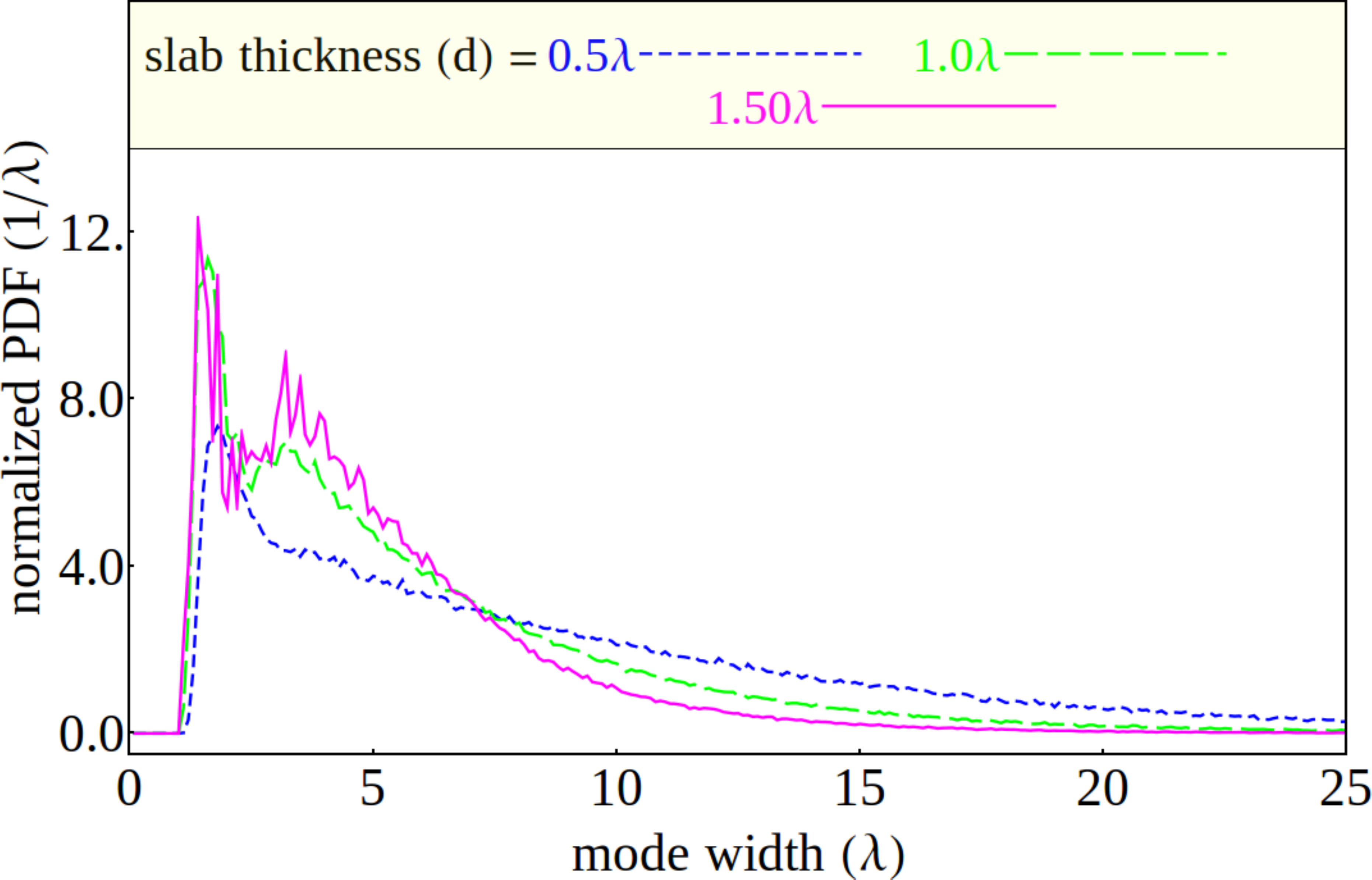}
}
\caption{\label{fig:FWO1}Normalized mode-width PDF of disordered waveguides defined by ${\Delta n_{\rm core}} = 0.01$, ${\Delta n_{\rm clad}} = 0$, $N=200$, and unit slab thickness of $d=0.5, 1.0,$ and $1.5\lambda$. Overall, the waveguides show stronger localization for a thicker unit slab.}
\centerline{
\includegraphics[width=1.0\columnwidth]{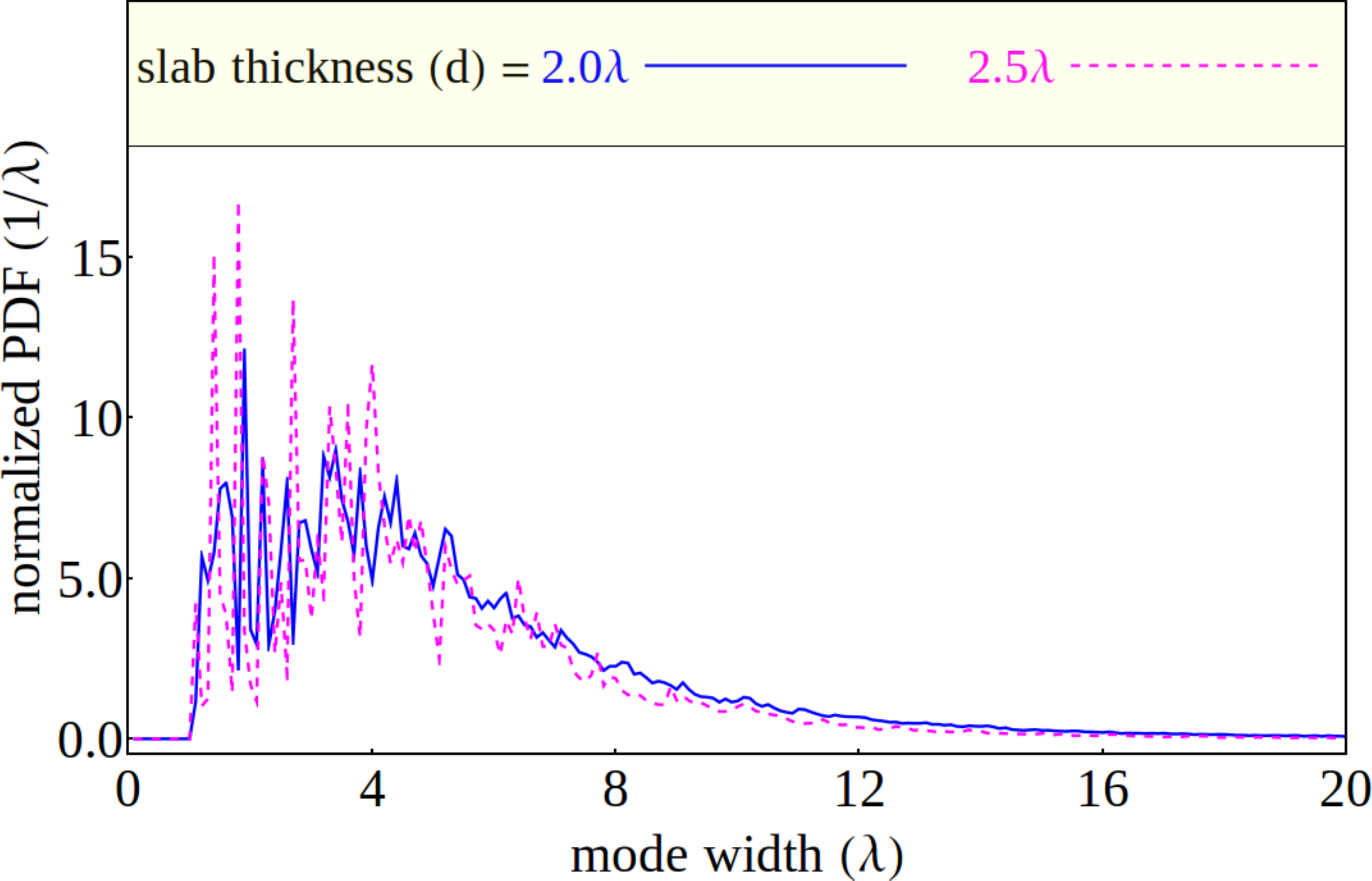}
}
\caption{\label{fig:FWO2}The same as Fig.~\ref{fig:FWO1}, except that the thickness of the unit slabs in the structure of the disordered waveguides is larger. The formation of local waveguides leads to sharp discrete peaks in the normalized PDF.}
\end{figure}

%%%%%%%%%%%%%%%%%%%%%%%%%%%%%%%%%%%%%%%%%%%%%%%%%%%%%%%%%%%%%%%%%%%%%%%%%%%
In the previous sections, we learned much about the behavior of the mode-width PDF for various
refractive index configurations in the core and cladding. In all previous simulations,
we assumed that the width of each slab is equal to the wavelength $d=\lambda$. However,
the mode-width PDF depends on the value of $d$ as well. Understanding the behavior of 
the mode-width PDF as a function of $d$ is quite important because $d$ is a parameter
that can be used to optimize the disordered lattice given an objective function. 
For example, our objective can be to obtain the smallest possible mean value of the 
mode width calculated using the PDF, where $d$ in addition to the refractive indexes
can be used as an optimization parameter.
In Figs.~\ref{fig:FWO1} and~\ref{fig:FWO2} we plot the normalized PDFs for disordered 
lattices defined by $n_{c} = n_{0}=1.49$, and $n_1=1.50$.
The value of the unit slab 
thickness is different in each case, taking the values ranging over $d=0.5\lambda-2.5\lambda$,
while keeping the total waveguide width equal to $200\lambda$. Therefore, the case with
$d=0.5\lambda$ corresponds to $N=400$, while the case with $d=2\lambda$ corresponds to $N=100$.
As we discussed before, the
normalized PDF integrates to the total number of modes, which varies from an average of
49 modes for the case of $d=0.5\lambda$ to an average of 39 modes for the case of $d=2.5\lambda$.
In Fig.~\ref{fig:FWO1}, it is clear that $d=0.5\lambda$ corresponds to a normalized mode-width PDF 
with a long tail in the extended mode region. When $d$ is increased to $d=\lambda$ and
further to $d=1.5\lambda$, the extended tail is gradually lowered contributing more to 
the localized region. It seems as if that the extended modes trade off their role 
with the localized modes of the second PDF hump. Another important observation is that the localized peak 
shifts slightly towards the smaller mode width values as the unit slab thickness increases.
%%%%%%%%%%%%%%%%%%%%%%%%%
%%%%%%%%%%%%%%%%%%%%%%%%%
%%%%%%%%%%%%%%%%%%%%%%%%%
%%%%%%%%%%%%%%%%%%%%%%%%%   

In Fig.~\ref{fig:FWO2}, the normalized PDFs for disordered lattices with $d=2\lambda$ and
$d=2.5\lambda$ are shown. The normalized PDF in these figures exhibit sharp peaks, which are markedly
different from the PDFs we have observed in previous figures. Below, we will argue that these sharp 
peaks are mainly due to step-index waveguiding behavior of individual discrete waveguides accidentally formed
in the random structure. In order to understand this, 
consider the case of $d=2\lambda$, where discrete local waveguides of widths $t=2\lambda$,
$t=4\lambda$, $t=6\lambda$, etc appear, respectively, with decreasing probability. The V-number of the slab 
waveguide from Eq.~\ref{eq:V-number} ($n_{\rm co}=n_1$ and $n_{\rm cl}=n_0$) is
equal to 1.09 for $t=2\lambda$ and is proportionally larger for $t=4\lambda$, $t=6\lambda$, etc.
We recall that the single-mode cut-off condition for the TE modes of a slab waveguide is $V=\pi/2$.
Therefore, $t=2\lambda$ is near cut-off and $t=4\lambda$ or larger are multimode. The large
V-number in these waveguides results in highly confined modes that cannot interact with the modes of the
neighboring waveguides to allow for randomized interaction to form Anderson localized modes. Therefore,
in addition to the extended modes and the Anderson localized modes that stem from the more-loosely-bound modes,
we encounter the regular step-index waveguiding modes in the form of sharp discrete peaks. The peaks are centered 
at mode-width values of the corresponding waveguides of discrete thickness values of $t=2\lambda$,
$t=4\lambda$, $t=6\lambda$, etc. The decreasing values of the discrete peaks in the PDF are indicative of the 
decreasing probability of having local waveguides with $t=2\lambda$,
$t=4\lambda$, $t=6\lambda$, etc, respectively. This situation is even more prominent in the case of
$d=2.5\lambda$, where discrete local waveguides have widths of $t=2.5\lambda$,
$t=5\lambda$, $t=7.5\lambda$, etc.

\begin{figure}[t]
\centerline{
\includegraphics[width=1.0\columnwidth]{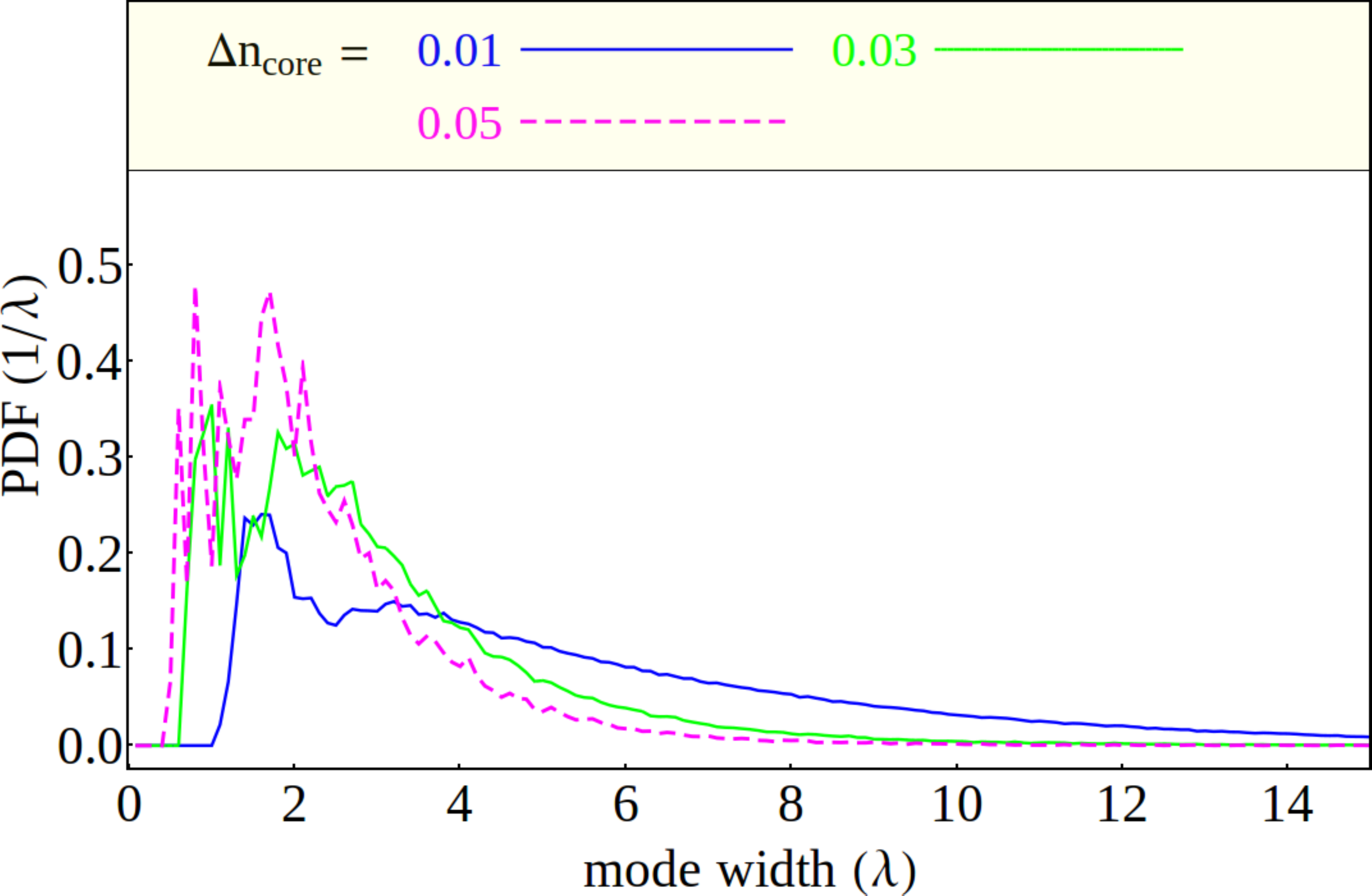}
}
\caption{\label{fig:increase-deltan}The unit slab thickness is fixed at $d = \lambda$ but ${\Delta n_{\rm core}} = 0.01, 0.03,$ and $0.05$. As ${\Delta n_{\rm core}}$ increases, more local waveguides form in the structure and the PDF becomes more discrete.}
\end{figure}

Our argument in the previous paragraph was based on the value of the V-number created in the locally formed waveguides. Therefore,
if the refractive index contrast ${\Delta n_{\rm core}}$ is increased, we should observe a similar behavior, where narrow peaks related to  
regular step-index waveguiding modes should appear alongside with the extended modes and the Anderson localized modes.
The discrete peaks in the PDF observed in Fig.~\ref{fig:PDFcase2} are in fact of this nature. In order to see this more clearly, 
in Fig.~\ref{fig:increase-deltan} we study the impact of increasing the value of the waveguide index difference ${\Delta n_{\rm core}}$ 
by comparing the mode-width PDFs for ${\Delta n_{\rm core}=0.01}$, ${\Delta n_{\rm core}=0.03}$, and ${\Delta n_{\rm core}=0.05}$,
all for $d=\lambda$. Sharp peaks clearly appear when the refractive index contrast is increased.
%%%%%%%%%%%%%%%%%%%%%%%%%
%%%%%%%%%%%%%%%%%%%%%%%%%
%%%%%%%%%%%%%%%%%%%%%%%%%
%%%%%%%%%%%%%%%%%%%%%%%%%

The results presented in this section so far give a thorough overview on the statistical behavior of Anderson localized modes, extended modes, and regular 
step-index waveguiding modes, all of which can be present in a disordered waveguide at the same time.
%%%%%%%%%%%%%%%%%%%%%%%%%%%%%%%%%%%%%%%%%%%%%%%%%%%%%%%%%%%%%%%%%%%%%%%%%%%%%%%%%%%%%%%%%%%%%%%%
%%%%%%%%%%%%%%%%%%%%%%%%%%%%%%%%%%%%%%%%%%%%%%%%%%%%%%%%%%%%%%%%%%%%%%%%%%%%%%%%%%%%%%%%%%%%%%%%

%%%%%%%%%%%%%%%%%%%%%%%%%
%%%%%%%%%%%%%%%%%%%%%%%%%
\begin{figure}[htp]
\centerline{
\includegraphics[width=0.9\columnwidth]{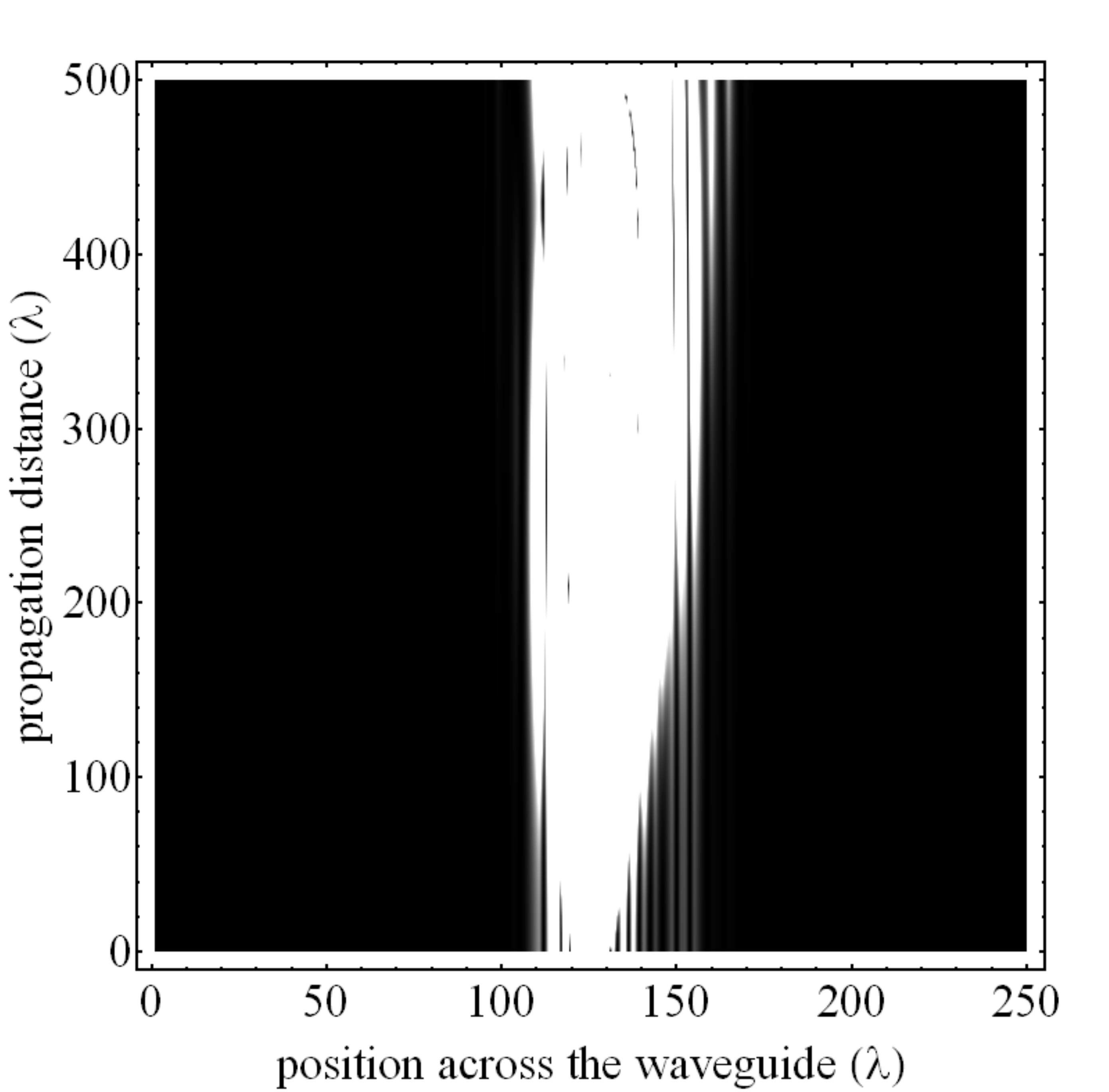}
}
\caption{\label{fig:MD1}Propagation of a Gaussian beam ($\omega = 3\lambda$) along a disordered waveguide defined by ${\Delta n_{\rm core}} = 0.01$, ${\Delta n_{\rm clad}} = 0,$ and $N=200$. The
beam eventually localizes to a relatively stable width after an initial expansion.}
\centerline{
\includegraphics[width=0.9\columnwidth]{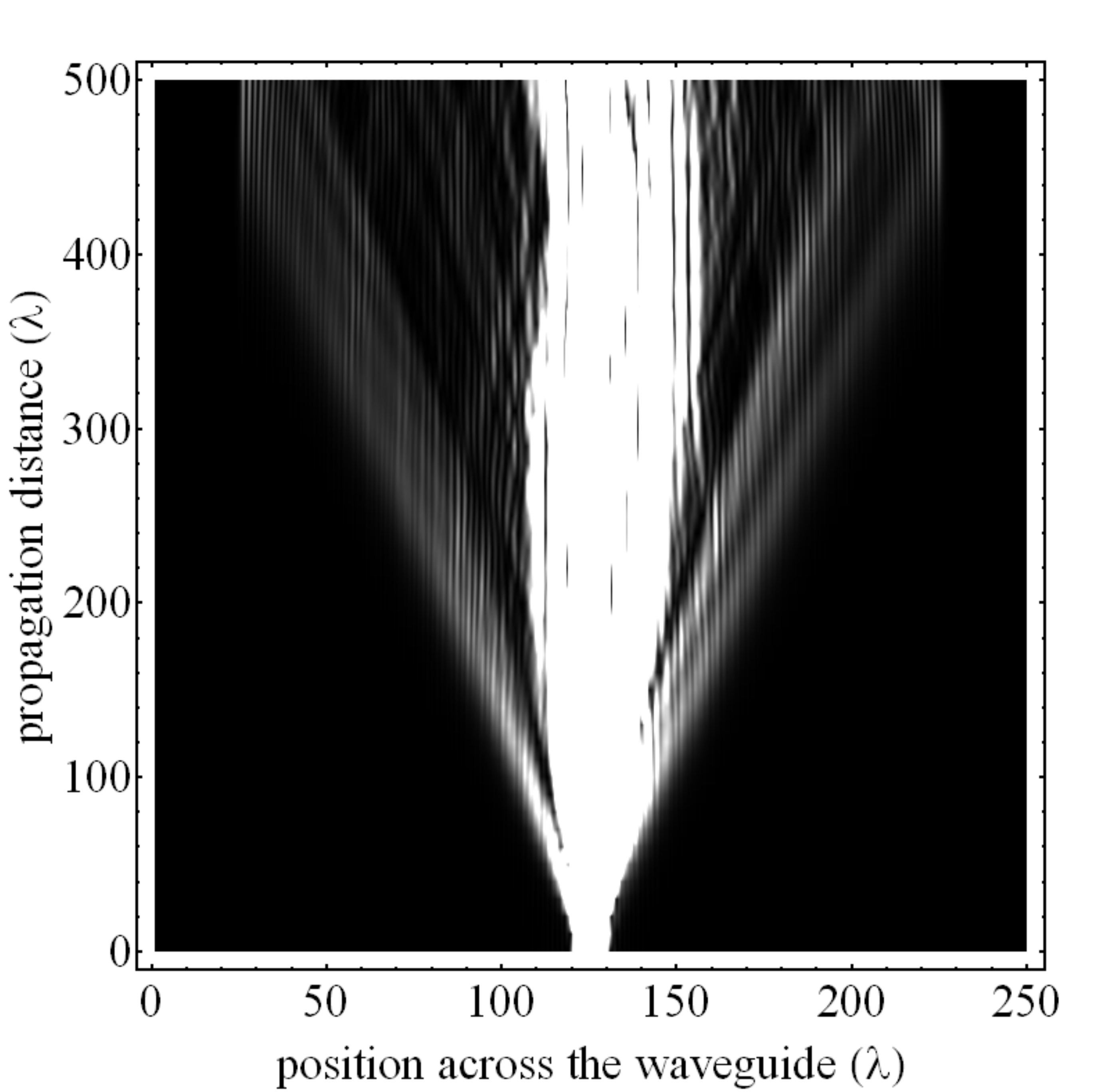}
}
\caption{\label{fig:MD2}The refractive index profile of the disordered waveguide is exactly the same as Fig.~\ref{fig:MD1}, except ${\Delta n_{\rm clad}} = 0.04$. Formation of the extended modes generates a background noise.}
\end{figure}
%%%%%%%%%%%%%%%%%%%%%%%%%
%%%%%%%%%%%%%%%%%%%%%%%%% 
We conclude this section by visualizing the interplay between the impact of the localized and 
extended modes in a disordered waveguide. In Fig.~\ref{fig:MD1}, we numerically simulate the 
propagation of light in a disordered waveguide and plot the intensity distribution of the guided 
beam as it propagates along the waveguide. The disordered waveguide is defined with $N=200$ slabs, 
where each slab's thickness is $d=\lambda$, and the refractive indexes are given by $n_{1}$ = 1.5 
and $n_{0}$ = 1.49. The cladding index in Fig.~\ref{fig:MD1} is $n_{c}$ = 1.49, so ${\Delta n_{\rm clad}=0}$,
while the cladding index in Fig.~\ref{fig:MD2} is $n_{c}$ = 1.45 resulting in ${\Delta n_{\rm clad}=0.04}$.
Based on the results of the previous section and the discussion on the normalized mode-width PDF, 
we expect to have nearly identical distribution of localized modes. However, the larger 
value of ${\Delta n_{\rm clad}}$ in Fig.~\ref{fig:MD2} results in a larger number of extended modes.

In Figs.~\ref{fig:MD1} and \ref{fig:MD2}, the injected beam is a Gaussian characterized by the electric field distribution 
of the form $E(x) \propto \exp(-x^2/\omega^2)$ with $\omega = 3\lambda$ at the entrance, where $x$ is the coordinate 
across of the waveguide. The center of the Gaussian beam is assumed to be in the middle of the disordered lattice. 
In the single realization of the disordered waveguide shown in Fig.~\ref{fig:MD1} with very few extended modes, 
there is virtually no background noise and the initial excitation is clearly Anderson localized after a short propagation
distance. However, in the presence of a large number of extended modes in Fig.~\ref{fig:MD2}, a background noise 
due to extended modes is evident throughout the propagation, while the Anderson localized modes still play a prominent 
role in the center that is similar to the one observed in Fig.~\ref{fig:MD1}.
%%%%%%%%%%%%%%%%%%%%%%%%%%%%%%%%%%%%%%%%%%%%%%%%%%%%%%%%%%%%%%%%%%%%%%%
%%%%%%%%%%%%%%%%%%%%%%%%%%%%%%%%%%%%%%%%%%%%%%%%%%%%%%%%%%%%%%%%%%%%%%%
\section{Discussion}
%%%%%%%%%%%%%%%%%%%%%%%%%%%%%%%%%%%%%%%%%%%%%%%%%%%%%%%%%%%%%%%%%%%%%%%
There is a vast literature over the past five decades on Anderson localization, especially in 1D, which is the 
main focus of this paper. In this section, we will establish a connection between the key aspects of the work presented here 
and the existing literature. In particular, scaling properties of electron transport and conductance have received 
considerable attention over the years. There is a one-to-one relationship between the 
Schr\"{o}dinger equation for electron in a disordered potential $V(x)$
%%%%%%%%%%%%
\begin{align}
\label{eq:schrodinger}
-\dfrac{-\hbar^2}{2m}\dfrac{\partial^2}{\partial x^2}A(x)+V(x)A(x)=EA(x),
\end{align}
%%%%%%%%%%%%
and the paraxial Helmholtz equation~\ref{eq:helmholtz} for optical wave propagation--along the z-direction--in a 
longitudinally (z-direction) invariant and transversally (x-direction) disordered waveguide. The analogy can be 
established by making the following identifications:
%%%%%%%%%%%%
\begin{subequations}
\begin{align}
V(x)&=\dfrac{\hbar^2k_0^2}{2m}\Big(n_{\rm cl}^2-n^2(x)\Big),\\
E&=\dfrac{\hbar^2k_0^2}{2m}\Big(n_{\rm cl}^2-n^2_{\rm eff}\Big),
\end{align}
\label{eq:schrodinger2}
\end{subequations}
%%%%%%%%%%%%
where $n_{\rm eff}=\beta/k_0$. Our study in this manuscript has focused on guided waves with
$n_{\rm eff}>n_{\rm cl}$, which is equivalent to the problem of electronic bound-states 
with $E<0$ in Eq.~\ref{eq:schrodinger} (in the outer left and right boundaries we 
have $V(x\in{\rm boundary})=0$).
In Ref.~\cite{SalmanModal}, we established a relationship between the mode-width of the localized 
states and the localization length. Briefly, for an exponentially localized state of the form $A(x)\sim\exp(-|x|/L_c)$,
the mode-width ($\sigma$ defined by Eq.~\ref{eq:sigma}) is given by $\sigma=\sqrt{2}L_c$. The localization 
length ${\bar L}_c$, is defined through logarithmic averaging of the localized beam profile 
intensity (modulus-squared)~\cite{Soukoulis2}.

Scaling theories of localization have been discussed in multiple publications especially 
in late 1970s and early 
1980s~\cite{THOULESS197493,Wegner1976,Licciardello,Abrahams,Anderson1980,Stone,Pichard1981-1,Pichard1981-2,Pichard1986-1,Pichard1986-2}.
These and similar work have primarily focused on the scaling properties of conductivity. There are similarities between 
the scaling analyses of these papers and our work especially at the formal level of the governing differential equations~\ref{eq:helmholtz}
and~\ref{eq:schrodinger}. However, there are subtle and important differences which arise primarily due the physical nature of the problem
here which only deals with the transversely localized guided optical modes propagating in the longitudinal direction. The differences 
are mathematically manifested in the different boundary conditions used in these problems. 
For example, consider
the work of Pichard~\cite{Pichard1986-1} that studies the 1D scaling of the Anderson model and resembles our work
because of the 1D nature of both problems and the underlying Eqs.~\ref{eq:schrodinger} and~\ref{eq:schrodinger2}. Pichard
studies the scaling behavior of the eigenvalue $\lambda_N$ of the unimodular matrix ${\tau}^\dagger_N{\tau}_N$ with the 
length of a disordered chain $N$, where ${\tau}_N$ is the $2\times 2$ transfer matrix of the 1D disordered system. 
For the analysis, Pichard emphasizes that the boundary condition is that of electronic waves of the Fermi surface with a real 
wave-vector. Therefore, Pichard studies the scattering problem using Eq.~\ref{eq:schrodinger} with $E>0$ 
(for $V(x\in{\rm boundary})=0$) and explores the scaling behavior of $\lambda_N$ where both $\lambda_N$ and resistance depend on the value of $E$. 
In this manuscript, we formally study the same differential equation~\ref{eq:schrodinger} as that of Pichard, with the minor 
difference that Ref.~\cite{Pichard1986-1} only considers diagonal disorder and our disorder is mixed due to the practical nature of the
studied problem. However, the main difference arises in the boundary condition, where we treat Eq.~\ref{eq:schrodinger} as an eigenvalue problem
and only study bound-states with $E<0$, hence exponentially decaying tails of the eigenstates because $V(x\in{\rm boundary})=0$ (or equivalently 
an imaginary wave-vector in the boundary).
Moreover, we focus our studies on the mode-widths of these {\em eigenstates} which are roughly related to their near exponential decay 
(on each side) through $\sigma=\sqrt{2}L_c$ that was derived above (and there is no $E$ dependence because $E$ here is an eigenvalue which we solve for). 
Therefore, unlike Ref.~\cite{Pichard1986-1} that analyses the
{\em eigenvalues} of the modulus squared of the transmission matrix, our focus is on the {\em eigenstates} of Eq.~\ref{eq:schrodinger} with $E<0$ (bound-states). For each disordered 
waveguide, we obtain the mode-widths for a large number of bound-states. As expected, there is a (somewhat non-trivial) 
relationship between the decay exponent $L_c$ (and localization properties) of an eigenstate and the corresponding eigenvalue, which is 
discussed in Refs.~\cite{MafiAOP,Lahini}. Similarly, the PDF has been studied extensively in the past (see Refs.~\cite{disorderedBook,BeenakkerPRM}
and references therein), again in the context of the localization length determined through the scattering problem discussed above.
Our analysis is focused on the PDF of the eigenvalue problem and the emphasis has been placed on the statistics and scaling of the PDF of the
mode-width directly calculated from the eigenstates, which is the relevant quantity for the experiments presented on disordered optical fibers 
in references such as Ref.~\cite{SalmanOL} and Ref.~\cite{SalmanNature}. We would also like to acknowledge an interesting body of work on the scaling 
properties of the scattering problems in optical systems, e.g. in Ref.~\cite{Aegerter:07} which resemble more the work of electron transport~\cite{disorderedBook,BeenakkerPRM,Pendry} than 
the bound-state problem studied here.  
%%%%%%%%%%%%%%%%%%%%%%%%%%%%%%%%%%%%%%%%%%%%%%%%%%%%%%%%%%%%%%%%%%%%%%%
%%%%%%%%%%%%%%%%%%%%%%%%%%%%%%%%%%%%%%%%%%%%%%%%%%%%%%%%%%%%%%%%%%%%%%%
\section{conclusion}
%%%%%%%%%%%%%%%%%%%%%%%%%%%%%%%%%%%%%%%%%%%%%%%%%%%%%%%%%%%%%%%%%%%%%%%
In this manuscript, we have introduced the mode-width PDF as a powerful tool to study the
transverse Anderson localization properties of guided modes of a disordered one dimensional optical waveguide. 
The mode-width PDF has been used for detailed statistical analysis of the impact of various structural and 
optical parameters of the disordered waveguide. A disordered waveguide supports both Anderson localized modes
as well as extended modes. The mode-width PDF sheds light into the distribution of these modes and provides
a powerful framework to manipulate such distributions, for example to quench the number of extended 
modes while minimally affecting the localized ones.
An important observation in this manuscript is the convergence of the mode-width PDF to a terminal configuration
as a function of the transverse dimension of the disordered waveguide. This has been shown by performing a scaling analysis
of the mode-width PDF and can be quite helpful in turning a formidable computational problem from nearly impossible to a tractable one.
The results presented in the manuscript are intended to establish the framework for a comprehensive analysis of the mode-width statistics 
for 2D transverse Anderson localization in optical fibers in the future.    
%%%%%%%%%%%%%%%%%%%%%%%%%%%%%%%%%%%%%%%%%%%%%%%%%%%%%%%%%%%%%%%%%%%%%%%
%%%%%%%%%%%%%%%%%%%%%%%%%%%%%%%%%%%%%%%%%%%%%%%%%%%%%%%%%%%%%%%%%%%%%%%
\bibliography{refs}
\end{document}